\documentclass[onecolumn,a4paper,fleqn,usenatbib]{mnras}
\usepackage{newtxtext,newtxmath}

\usepackage[T1]{fontenc}
\usepackage{ae,aecompl}

\usepackage{graphicx}	

\title[Coagulation of icy dust aggregates]{Is water ice an efficient facilitator for dust coagulation?}

\author[H. Kimura et al.]{Hiroshi Kimura,$^{1}$\thanks{E-mail: hiroshi\_kimura@perc.it-chiba.ac.jp}
Koji Wada,$^{1}$
Hiroshi Kobayashi,$^{2}$
Hiroki Senshu,$^{1}$
Takayuki Hirai,$^{1}$
\newauthor
Fumi Yoshida,$^{1,3}$
Masanori Kobayashi,$^{1}$
Peng K. Hong,$^{1}$
Tomoko Arai,$^{1}$
Ko Ishibashi$^{1}$
\newauthor
and Manabu Yamada$^{1}$
\\
$^{1}$Planetary Exploration Research Center (PERC), Chiba Institute of Technology, 
Tsudanuma 2-17-1, Narashino, Chiba 275-0016, Japan\\
$^{2}$Department of Physics, Nagoya University, Chikusa-ku Furo-cho, Nagoya 464-8602, Japan\\
$^{3}$School of Medicine, Department of Basic Sciences, University of Occupational and Environmental Health, Japan, 1-1 Iseigaoka, Yahata, Kitakyusyu 807-8555, Japan\\
}

\date{Accepted 2020 August 12. Received 2020 August 6; in original form 2020 June 30}

\pubyear{2020}

\begin{document}
\volume{498}
\label{firstpage}
\pagerange{1801--1813}
\maketitle

\begin{abstract}
Beyond the snow line of protoplanetary discs and inside the dense core of molecular clouds, the temperature of gas is low enough for water vapour to condense into amorphous ices on the surface of preexisting refractory dust particles.
Recent numerical simulations and laboratory experiments suggest that condensation of the vapour promotes dust coagulation in such a cold region.
However, in the numerical simulations, cohesion of refractory materials is often underestimated, while in the laboratory experiments, water vapour collides with surfaces at more frequent intervals compared to the real conditions.
Therefore, to re-examine the role of water ice in dust coagulation, we carry out systematic investigation of available data on coagulation of water ice particles by making full use of appropriate theories in contact mechanics and tribology.
We find that the majority of experimental data are reasonably well explained by lubrication theories, owing to the presence of a quasi-liquid layer (QLL).
Only exceptions are the results of dynamic collisions between particles at low temperatures, which are, instead, consistent with the JKR theory, because QLLs are too thin to dissipate their kinetic energies.
By considering the vacuum conditions in protoplanetary discs and molecular clouds, the formation of amorphous water ice on the surface of refractory particles does not necessarily aid their collisional growth as currently expected.
While crystallisation of water ice around but outside the snow line eases coagulation of ice-coated particles, sublimation of water ice inside the snow line is deemed to facilitate coagulation of bare refractory particles.
\end{abstract}




\section{Introduction} \label{sec:intro}

Water ice is ubiquitous in the cold regions of the Universe, owing to the fact that hydrogen and oxygen are the two most abundant elements to form a solid such as icy dust particles and comets.
It is, therefore, commonly accepted that the essential component of dust particles and planetesimals in protoplanetary discs is water ice beyond the so-called snow line, at which the temperature of gas is low enough for water vapour to condense into ices \citep*[e.g.,][]{cyr-et-al1998}.
Reactive accretion of water ice from hydrogen and oxygen atoms on the surface of dust particles takes place in the dense core of molecular clouds where the growth of dust particles has been observed by scattering of stellar radiation \citep{steinacker-et-al2010}.
It is worthwhile noting that laboratory experiments on the coagulation growth of water ice particles have a long history outside astronomy and planetary science, since coagulation is observed in daily life and is a plausible route to the formation of snowflakes \citep*[e.g.,][]{faraday1860,hosler-et-al1957}.
Recent works on laboratory measurements of cohesion between crystalline water ice particles at vacuum conditions provided encouraging results that dust particles composed of water ice might be much more cohesive than previously believed \citep{gundlach-et-al2011,gundlach-blum2015,jongmanns-et-al2017}.
Form a theoretical point of view, \citet*{chokshi-et-al1993} demonstrated that the JKR theory of elastic contact formulated by \citet*{johnson-et-al1971} is a powerful tool for better understanding of dust coagulation.
Numerical simulations incorporating the JKR theory have shown that dust aggregates consisting of submicrometre-sized water ice particles proceed with coagulation growth even at a collision velocity of $50~\mathrm{m~s^{-1}}$ \citep{wada-et-al2009,wada-et-al2013}.
As a result, the majority of recent studies on dust coagulation and planetesimal formation assume that silicate aggregates are disrupted by mutual collision at a velocity of $v_\mathrm{disrupt} \sim 1~\mathrm{m~s^{-1}}$, but icy aggregates at $v_\mathrm{disrupt} \sim 10~\mathrm{m~s^{-1}}$ \citep*[e.g.,][]{birnstiel-et-al2010,vericel-gonzalez2019}.
Such a trendy assumption led \citet{drazkowska-alibert2017} to propose planetesimal formation by the ``traffic jam'' effect at the snow line, provided that sticky water ice particles grow faster and thus drift toward the central star faster than less-sticky bare silicate particles, implying that aggregates of the former catch up the latter at the snow line, which results in a traffic jam.
However, we argue that the importance of water ice to dust coagulation is still open to debate, since water ice is not necessarily stickier than other materials such as silicates and complex organic matter \citep{kimura-et-al2015,kimura-et-al2020a,musiolik-wurm2019}.

Laboratory experiments on cohesion of water ice particles have been carried out at low-to-medium vacuum ($> 1~\mathrm{Pa}$) or atmospheric conditions ($\sim 10^{5}~\mathrm{Pa}$) up to date.
Note that even if the pressure is kept as low as $10^{-4}~\mathrm{Pa}$ in a vacuum chamber, water vapour is still present in the chamber and merely one second of time is sufficient for the vapour to form a monolayer on the surface of water ice particles.
Since water vapour more frequently collide with the surface of water ice particles at low and medium vacuum conditions than in protoplanetary discs and molecular clouds, successive condensation and evaporation of water vapour in the laboratory might significantly affect the experimental results.
Moreover, previous laboratory experiments were conducted in the range of temperatures where the surface of crystalline water ice is partly melted and covered by a thin quasi-liquid water film, referred to as a quasi-liquid layer (QLL)\footnote{The presence of QLLs on the surface of water ice was originally postulated by \citet{faraday1933}, but later it was experimentally confirmed and has now been widely accepted \citep*[e.g.,][]{kouchi-et-al1987,murata-et-al2016}.} \citep*{conde-et-al2008,kajima-et-al2014,slater-michaelides2019}.
In one of the early works\footnote{Throughout the paper, we use the word ``early works'' to differentiates the works that were done in the mid-20th century from ``recent works'' that were done in the beginning of the 21st century.} on {\it in situ} measurements of cohesion between water ice particles in the laboratory, \citet{nakaya-matsumoto1954} observed a rotation of the particles before separation, which was accounted for by the presence of QLLs on the surface of the particles.
Therefore, lubrication due to QLLs may play a vital role in laboratory experiments with water ice even at low-to-medium vacuum conditions, in spite of their negligible roles in protoplanetary discs and molecular clouds.
This signifies the importance of ultra-to-extremely high vacuum and low temperature conditions for laboratory experiments on a study of ices and its applications to astronomy and planetary science \citep{kouchi-et-al2018}.

By looking into previous experimental results in detail, most of the results with water ice indeed do not seem to be in accord with the JKR theory of contact mechanics.
Therefore, we re-examine laboratory experiments on the mechanical properties of water ice particles by making full use of currently available lubrication theories in tribology as a replacement for the JKR theory.
From a theoretical perspective, we will discuss as to whether water ice is an efficient facilitator of dust coagulation beyond the snow line in protoplanetary discs and in the dense core of molecular clouds. 

\section{Theoretical backgrounds}

\subsection{Contact mechanics}
\label{sec:JKR}

The maximum relative velocity between two colliding particles to proceed with dust coagulation is hereafter referred to as the critical velocity of sticking, $v_\mathrm{stick}$.
According to the JKR theory of contact mechanics, the critical velocity of sticking between two identical particles of radius $r_0$ and density $\rho$ is given by \citep{chokshi-et-al1993}
\begin{eqnarray}
v_\mathrm{stick} = \left({\frac{27 c_1 \mathrm{\upi}^{2/3}}{4}}\right)^{1/2} \left[{\frac{\gamma^5 \left({1 - \nu^2}\right)^2}{r_0^5 \rho^3 E^2}}\right]^{1/6} ,
\label{eq:v_stick}
\end{eqnarray}
where $\gamma$, $E$ and $\nu$ denote the surface energy, Young's modulus and Poisson's ratio, respectively.
Here $c_1$ is a constant on the order of unity and, according to \citet{chokshi-et-al1993}, we adopt a value of $c_1 = 0.935$.

Numerical simulations on mutual collision of dust aggregates provide an empirical formula for the critical velocity of disruption, $v_\mathrm{disrupt}$, above which the collisional velocity is too fast to promote coagulation growth against disruption \citep{wada-et-al2013,kimura-et-al2015}:
\begin{eqnarray}
v_\mathrm{disrupt} = c_2 \left({1.54 \frac{27 \mathrm{\upi}^{2/3}}{4}}\right)^{1/2} \left[{\frac{\gamma^5 \left({1-\nu^2}\right)^2}{r_0^5 \rho^3 E^2}}\right]^{1/6} ,
\label{eq:v_disrupt}
\end{eqnarray}
where $c_2$ is a constant in the range of $5.2$--$10$.

Table~\ref{tbl-1} lists the elastic parameters for crystalline and amorphous phases of water ice and silica, complex organic matter and amorphous carbon.
These materials are used as analogous to ices, silicates and carbonaceous matter that comprise dust particles in protoplanetary discs and molecular clouds.
We consider homogeneous dust particles and aggregates of the particles to re-examine experimental results with the use of pure water ice, although recent experimental works point to the presence of heterogeneity on the surface of dust particles \citep{rosufinsen-et-al2016,marchione-et-al2019}.
Also given in Table~\ref{tbl-1} are the critical velocities of sticking, $v_\mathrm{stick}$ and disruption, $v_\mathrm{disrupt}$, computed by equations~(\ref{eq:v_stick}) and (\ref{eq:v_disrupt}), respectively.
\begin{table}
 \centering
 \caption{Elastic properties and critical velocities.}
 \label{tbl-1}
 \begin{tabular}{lccccccl}
  \hline
Composition & $\rho$ & $\gamma$ & $E$ & $\nu$ & $v_\mathrm{stick}$$^\dagger$ & $v_\mathrm{disrupt}$$^\dagger$ & Reference\\ 
 & ($\mathrm{10^3~kg~m^{-3}}$) & ($\mathrm{J~m^{-2}}$) & ($\mathrm{GPa}$) &  & ($\mathrm{m~s^{-1}}$) & ($\mathrm{m~s^{-1}}$) &  \\
  \hline
c-H$_2$O  & 1.0 & 0.24 & 7  & 0.25 &  $1.2 \times {10}^1$  &  $(0.8 - 1.6) \times {10}^2$ & \citet{wada-et-al2007,pan-et-al2010}\\
a-H$_2$O  & 1.0 & 0.11 & 7  & 0.25 &  $6.7 \times {10}^0$  &  $(4.4 - 8.5) \times {10}^1$ & \citet{wada-et-al2007,kimura-et-al2020a}\\
c-SiO$_2$  & 2.0 & 1.5 & 70 & 0.17 &  $1.9 \times {10}^1$  & $(1.3 - 2.4) \times {10}^2$ & \citet{kimura-et-al2020a,kimura-et-al2020b} \\
a-SiO$_2$  & 2.0 & 0.24 & 70 & 0.17 &  $4.1 \times {10}^0$  &  $(2.8 - 5.3) \times {10}^1$ & \citet{kimura-et-al2020a,kimura-et-al2015} \\
CHON            & 1.2 & 0.073 & 0.00012 & 0.48 &  $1.5 \times {10}^2$  &  $(1.0 - 1.9) \times {10}^3$ & \citet*{potschke-et-al2002,mcnicholas-rankilor1969} \\
a-C         & 1.7 & 0.034 & 120 & 0.30 &  $0.7 \times {10}^0$  &  $(4.8 - 9.2) \times {10}^0$ & \citet{piazza-morell2009,marques-et-al2003} \\
  \hline
\multicolumn{8}{l}{$^\dagger$The values of $v_\mathrm{stick}$ and $v_\mathrm{disrupt}$ are estimated for particles of radius $r_0 = 0.1~\micron$ and aggregates of these particles, respectively}\\
 \end{tabular}
\end{table}

In the framework of the JKR theory, the rolling friction force $F_\mathrm{roll}$ is known to be independent of particle radius, as given by \citep{dominik-tielens1995,dominik-tielens1997}
\begin{eqnarray}
F_\mathrm{roll} & = & 6 \mathrm{\upi} \gamma \xi_\mathrm{crit} ,
\label{eq:froll_JKR}
\end{eqnarray}
where $\xi_\mathrm{crit}$ is the critical displacement, a typical value of which is $\xi_\mathrm{crit} = 0.2~\mathrm{nm}$, because it should be of the order of the distance between neighbouring atoms \citep{dominik-tielens1995}.

In the JKR theory, the critical force $F_\mathrm{pull}$ required to pull off a particle of radius $r_1$ from a particle of radius $r_2$ is given by \citet{johnson-et-al1971}:
\begin{eqnarray}
F_\mathrm{pull} &=& 3 \mathrm{\upi} \gamma R, 
\label{eq:fpull_JKR}
\end{eqnarray}
with the reduced radius $R \equiv r_1 r_2 / (r_1 + r_2)$.

\citet{dominik-tielens1996,dominik-tielens1997} give an analytic formula for the sliding friction force $F_\mathrm{slide}$ based on the JKR theory:
\begin{eqnarray}
F_\mathrm{slide} & = & \left({\frac{9 \gamma \left({1-\nu^2}\right) R^2}{2^{7/2} \mathrm{\upi}^{1/2} E}}\right)^{2/3} \left[{\frac{E}{1 + \nu}  - 1.78 \mathrm{\upi} \left({\frac{b}{a}}\right)^3 \frac{E}{1 + \nu} + 65.92 \mathrm{\upi} \left({\frac{b}{a}}\right)^4 \frac{\gamma}{a}}\right] ,
\label{eq:fslide_JKR}
\end{eqnarray}
where $a$ and $b$ are material dependent constants ($a = b = 0.336~\mathrm{nm}$ for water ice).

\subsection{Tribology}
\label{sec:capillary}

\begin{figure}
\centering\includegraphics[width=0.4\columnwidth]{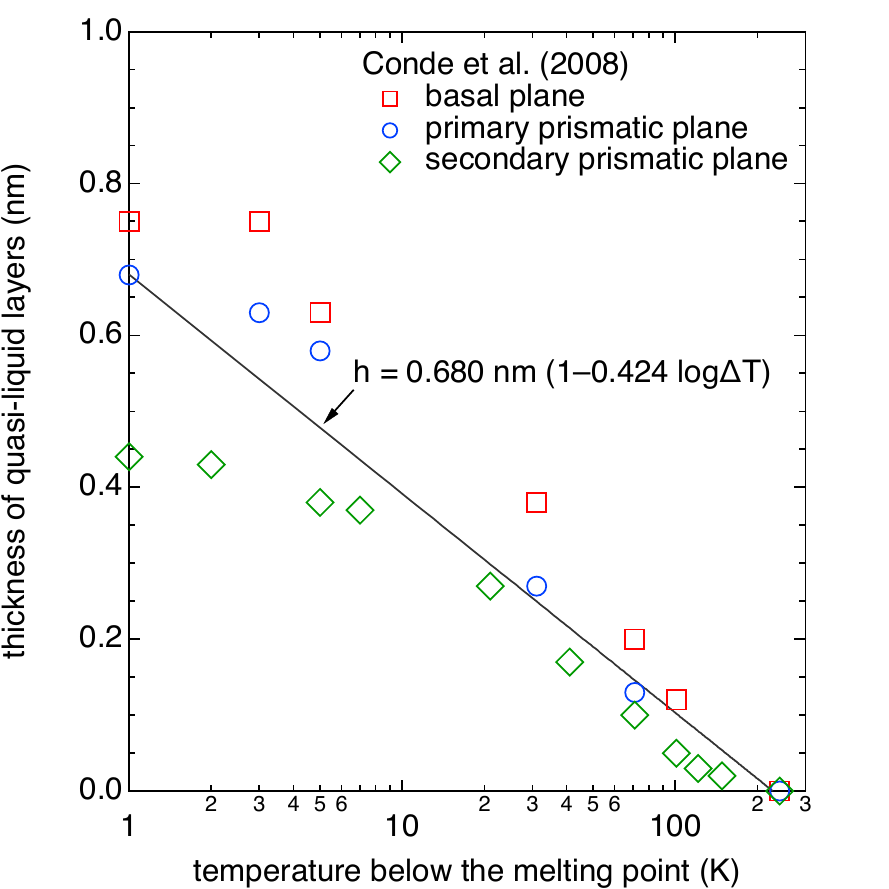}
\caption{The thickness $h$ of quasi-liquid layers as a function of the temperature $T$ below the melting point $T_\mathrm{m}$. Molecular dynamics simulation by \citet{conde-et-al2008}: the basal plane (squares), the primary prismatic plane (circles) and the secondary prismatic plane (diamonds); solid line: $h = 0.680~\mathrm{nm} \left({1 - 0.424\, \log \Delta T}\right) $ with $\Delta T = T_\mathrm{m} - T$.
\label{fig:h_QLL}}
\end{figure}
Provided that the surfaces of contacting water ice particles are covered by QLLs with thickness $h$ and surface tension $\sigma $, the particles adhere each other by capillary force.
According to \citet{zakerin-et-al2013}, the pull-off force $F_\mathrm{pull}$ of water ice particles with QLLs, which equals to the capillary force, is given by
\begin{eqnarray}
F_\mathrm{pull} & = & 4 \mathrm{\upi} \sigma R + \frac{\mathrm{\upi}^3}{3} \left({\frac{1-\nu^2}{E}}\right)^2 R^2 \left({\frac{\sigma}{r_\mathrm{K}}}\right)^3 + 2 \mathrm{\upi} h R \frac{\sigma}{r_\mathrm{K}} ,
\label{eq:fadh}
\end{eqnarray}
with the radius $r_\mathrm{K}$ of curvature for meniscus, referred to as the Kelvin radius.
The ratio $\sigma/r_\mathrm{K}$ of surface tension to the Kelvin radius is determined by the Kelvin equation:
\begin{eqnarray}
\frac{\sigma }{r_\mathrm{K}} &=& - \frac{k_\mathrm{B} T }{V_\mathrm{m}} \ln{\phi},
\end{eqnarray}
where $\phi$ denotes the relative humidity, namely, the ratio of partial vapour pressure $p_\mathrm{v}$ to saturated vapour pressure $p_\mathrm{sat}$ at temperature $T$, $V_\mathrm{m}$ is the volume of a water molecule and $k_\mathrm{B}$ is the Boltzmann constant.
According to \citet{doeppenschmidt-butt2000}, the thickness $h$ of QLLs at a temperature below the melting point, $T_\mathrm{m}$, may be approximated to
\begin{eqnarray}
h & = &  \alpha \left[{1 - \beta \log\left({T_\mathrm{m} - T}\right)}\right] ,
\end{eqnarray}
with two fitting constants of $\alpha$ and $\beta$.
\citet{conde-et-al2008} derived the thickness $h$ of QLLs on the surface of ice Ih in the basal, the primary prismatic and the secondary prismatic planes from their molecular dynamics (MD) simulations.
By fitting the results of their MD simulations as shown in Fig.~\ref{fig:h_QLL}, we may adopt $\alpha = 0.680~\mathrm{nm}$ and $\beta = 0.424$, although the thickness of QLLs is still open to debate \citep{slater-michaelides2019}.
We should mention that this is a conservative estimate for the thickness of QLLs in comparison with $\alpha = 32~\mathrm{nm}$ and $\beta = 0.65625$ proposed by \citet{doeppenschmidt-butt2000}.
The dependence of surface tension $\sigma$ on the temperature is given by
\begin{eqnarray}
\sigma &=& \gamma_\mathrm{a} - T \left({\frac{d\,\sigma}{d\,T}}\right) ,
\end{eqnarray}
where $\gamma_\mathrm{a}$ is the surface energy of a solid in amorphous phase.
We assume $\gamma_\mathrm{a} = 0.114~\mathrm{J~m^{-2}}$ and $d\sigma / dT = 0.142~\mathrm{mN~m^{-1}~K^{-1}}$ by extrapolating currently available experimental data on the surface tension of ordinary water to low temperatures \citep{kimura-et-al2020a}.

From low to medium vacuum, it is inevitable that evaporation and sorption of water molecules around the neck of contacting particles influence rolling friction forces \citep{butt-et-al2010,schade-marshall2011}.
Therefore, we may apply a theory for lubrication rolling friction forces to interpret experimental data on rolling friction forces on water ice particles at such a vacuum condition.
According to \citet{israelachvili2011}, the lubrication rolling force $F_\mathrm{roll}$ is given by
\begin{eqnarray}
F_\mathrm{roll} & = &  \frac{1}{5} \epsilon\left[{\frac{3 \left({1-\nu^2}\right)}{2E R^2}}\right]^{1/3} {F_\mathrm{pull}}^{4/3}   , 
\label{eq:froll-adh}
\end{eqnarray}
where $\epsilon$ is the fraction of energy dissipated during the friction.
By inserting equation~(\ref{eq:fadh}) into equation~(\ref{eq:froll-adh}), we obtain
\begin{eqnarray}
F_\mathrm{roll} & = & \frac{1}{5} \epsilon \left[{\frac{3 \left({1-\nu^2}\right)}{2E R^2}}\right]^{1/3} \left[{4 \mathrm{\upi} \sigma R + \frac{\mathrm{\upi}^3}{3} \left({\frac{\sigma}{r_\mathrm{K}}}\right)^3 \left({\frac{1-\nu^2}{E}}\right)^2 R^2 + 2 \mathrm{\upi} h \left({\frac{\sigma}{r_\mathrm{K}}}\right) R}\right]^{4/3} .
\label{eq:froll_cap_full}
\end{eqnarray}
The lubrication sliding force is given by \citep{israelachvili2011}
\begin{eqnarray}
F_\mathrm{slide} & = & \mu_\mathrm{s} F_\mathrm{pull} ,
\label{eq:fslide_cap}
\end{eqnarray}
where $\mu_\mathrm{s}$ is the sliding friction coefficient ($\mu_\mathrm{s} = 0.1$ for water ice).

\section{Interpretation of experimental data}

\subsection{Recent works in the beginning of the 21st century}

\subsubsection{\citet{gundlach-blum2015}}

\begin{figure}
\centering\includegraphics[width=0.4\columnwidth]{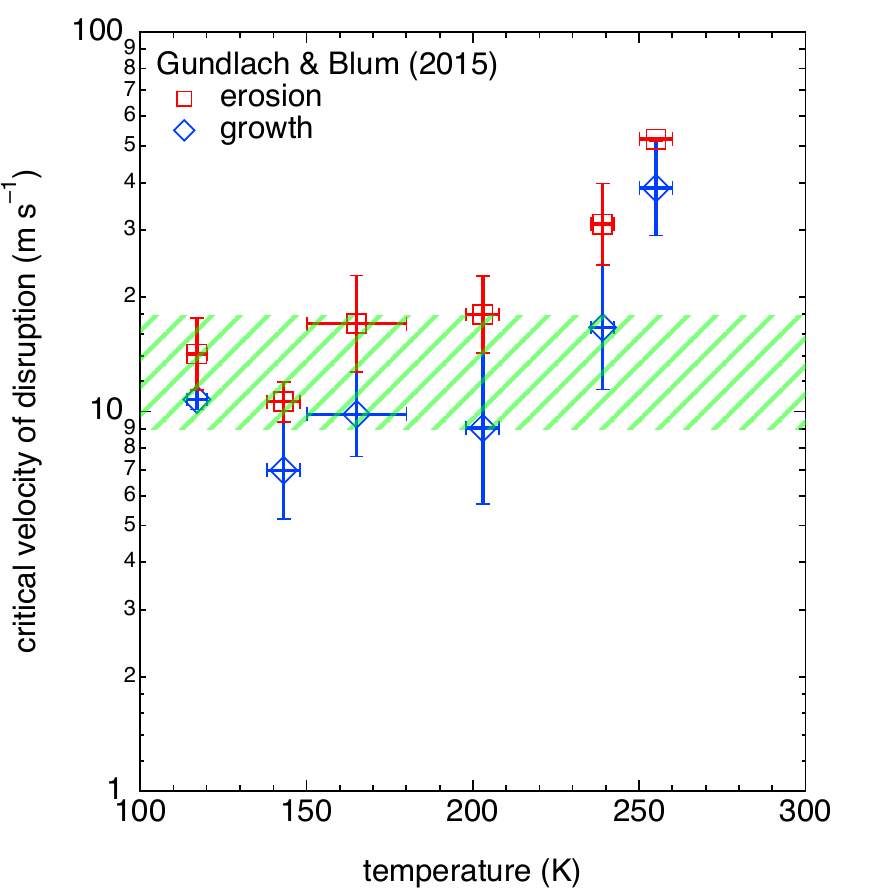}
\caption{Critical velocity of disruption $v_\mathrm{disrupt}$ for dust aggregates of micrometre-sized crystalline $\mathrm{H_2O}$ particles with radius $r_0 \approx 1.47~\micron$ as a function of temperature. Open diamonds and squares: experimental data \citep{gundlach-blum2015}; shaded area: equation~(\ref{eq:v_disrupt}) in the range of $c_2 = 5.2$--$10$.
\label{fig:v_disrupt}}
\end{figure}
In their laboratory experiments, \citet{gundlach-blum2015} imaged collisions between crystalline water ice particles of radius $r_0 \approx 1.47~\micron$ and an aggregate consisting of these particles.
The collisional velocity $v_\mathrm{imp}$ lay in the range of $v_\mathrm{imp} = 1$--$150~\mathrm{m~s^{-1}}$ and the temperature $T$ was controlled between $T \approx 114$ and $260~\mathrm{K}$ at a pressure of $\sim 100~\mathrm{Pa}$.
They observed the erosion of the aggregates at an impact velocity of $v_\mathrm{imp} \ga 15.3~\mathrm{m~s^{-1}}$ and the growth of the aggregates at an impact velocity of $v_\mathrm{imp} \la 9.6~\mathrm{m~s^{-1}}$.
The results were interpreted as convincing evidence that the presence of water ice helps the growth of dust aggregates, based on the presumption that the critical velocity $v_\mathrm{stick}$ of sticking between amorphous silica particles of the same radius is on the order of $v_\mathrm{stick} \sim 1~\mathrm{m~s^{-1}}$.
However, we would like to point out that there is a gap in logic here;
The critical velocity of sticking, $v_\mathrm{stick}$, is beside the point, because it does not correspond to the velocity that discriminates between the erosion and the growth of dust aggregates, but the critical velocity of disruption, $v_\mathrm{disrupt}$, does \citep[see][]{wada-et-al2009,wada-et-al2013}.
Accordingly, we compare the experimental results of \citet{gundlach-blum2015} to the critical velocity of disruption given by equation~(\ref{eq:v_disrupt}):
\begin{eqnarray}
v_\mathrm{disrupt} &=& 13.4 \pm 4.5~\mathrm{m~s^{-1}} \nonumber \\
&\times & \left({\frac{\gamma}{0.244~\mathrm{J~m^{-2}}}}\right)^{5/6} \left({\frac{r_0}{1.47~\micron}}\right)^{-5/6} \left({\frac{\rho}{1.0 \times {10}^{3}~\mathrm{kg~m^{-3}}}}\right)^{-1/2} \left({\frac{E}{7~\mathrm{GPa}}}\right)^{-1/3} \left[{\frac{1}{0.9375} - \frac{0.0625}{0.9375} \left({\frac{\nu}{0.25}}\right)^2}\right]^{1/3} . \nonumber \\
\label{eq:v_disrupt_ice}
\end{eqnarray}
As depicted in Fig.~\ref{fig:v_disrupt}, our estimates of $v_\mathrm{disrupt}$ are consistent with experimentally determined impact velocities at the boundary between the erosion and the growth of dust aggregates for the low temperatures $T < 239~\mathrm{K}$.
There are, however, noticeable deviations of experimental values from equation~(\ref{eq:v_disrupt_ice}) at $T \ga 239~\mathrm{K}$, which requires a mechanism of additional energy dissipation.
Using atomic force microscopy, \citet{doeppenschmidt-butt2000} measured the thickness of QLLs on the surface of water ice and concluded that the surface melting takes place at $T \ga 239~\mathrm{K}$.
Therefore, the increase of the experimental values at $T \ga 239~\mathrm{K}$ may be attributed to the efficient surface melting of water ice at $T \ga 239~\mathrm{K}$.

\subsubsection{\citet{gundlach-et-al2011}}

\begin{figure}
\centering\includegraphics[width=0.4\columnwidth]{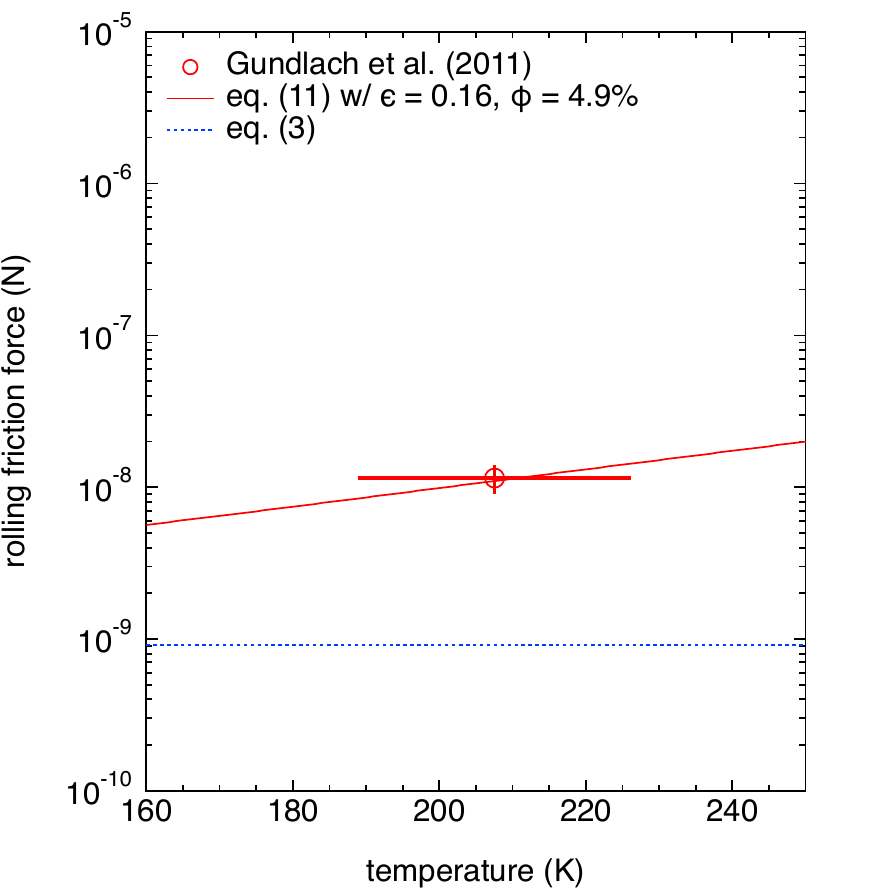}
\caption{Rolling friction forces of crystalline $\mathrm{H_2O}$ particles with radius $r_0 = 1.45~\micron$ as a function of temperature. Circle: experimental data \citep{gundlach-et-al2011}; solid line: equation~(\ref{eq:froll_cap_full}); dotted line: equation~(\ref{eq:froll_JKR}) or (\ref{eq:froll_ice}).
\label{fig:gundlach-et-al2011}}
\end{figure}
\citet{gundlach-et-al2011} measured the rolling friction force $F_\mathrm{roll}$ of porous dust aggregates consisting of micrometre-sized crystalline water ice particles with $r_0 = 1.45 \pm 0.65~\micron$, as well as aggregates of amorphous silica particles with $r_0 = 0.75~\micron$.
Rolling friction forces on the water ice aggregates were $F_\mathrm{roll} = (114.8 \pm 23.8) \times {10}^{-10}~\mathrm{N}$ in the range of temperatures from $189$ to $226~\mathrm{K}$ under nitrogen atmosphere.
If the JKR theory applies to their experiments, then we expect the rolling friction forces, according to equation~(\ref{eq:froll_JKR}), to be:
\begin{eqnarray}
F_\mathrm{roll} & = & 9.2 \times {10}^{-10}~\mathrm{N} \, \left({\frac{\gamma}{0.243~\mathrm{J~m^{-2}}}}\right) \left({\frac{\xi_\mathrm{crit}}{0.2~\mathrm{nm}}}\right) ,
\label{eq:froll_ice}
\end{eqnarray}
which is one order of magnitude smaller than the measured value.
\citet{gundlach-et-al2011} derived the surface energy of $\gamma = 0.19~\mathrm{J~m^{-2}}$ from equation~(\ref{eq:froll_JKR}) with $F_\mathrm{roll} = 114.8 \times {10}^{-10}~\mathrm{N}$ by assuming a critical displacement of $\xi_\mathrm{crit} = 3.2~\mathrm{nm}$.
We should, however, emphasize that an estimate of the surface energy in this manner strongly depends on the assumption of critical displacement, while the assumption of $\xi_\mathrm{crit} = 3.2~\mathrm{nm}$ has never been justified and the assumption of $\xi_\mathrm{crit} = 0.2~\mathrm{nm}$ has been shown to be consistent with their measurement of $F_\mathrm{roll}$ for amorphous silica \citep[see][]{kimura-et-al2015}.
Therefore, we regret that the surface energy of $\gamma = 0.19~\mathrm{J~m^{-2}}$ for water ice has been given very little credit and the application of the JKR theory to their experiments is thus in doubt.
Here, we shall interpret the experimental results of rolling friction forces measured by \citet{gundlach-et-al2011} in the framework of tribology, which is described in equation~(\ref{eq:froll_cap_full}).
Figure~\ref{fig:gundlach-et-al2011} demonstrates that the experimental data of \citet{gundlach-et-al2011} for water ice aggregates are well accounted for by lubrication, instead of the JKR theory, if $\epsilon = 0.16$ and $\phi = 4.9\%$ in equation~(\ref{eq:froll_cap_full}).
Therefore, we cannot help wondering if the rolling friction forces measured by \citet{gundlach-et-al2011} for crystalline water ice would be reduced by one order of magnitude at ultra-to-extremely high vacuum conditions.

\subsubsection{\citet{jongmanns-et-al2017}}

\begin{figure}
\centering\includegraphics[width=0.4\columnwidth]{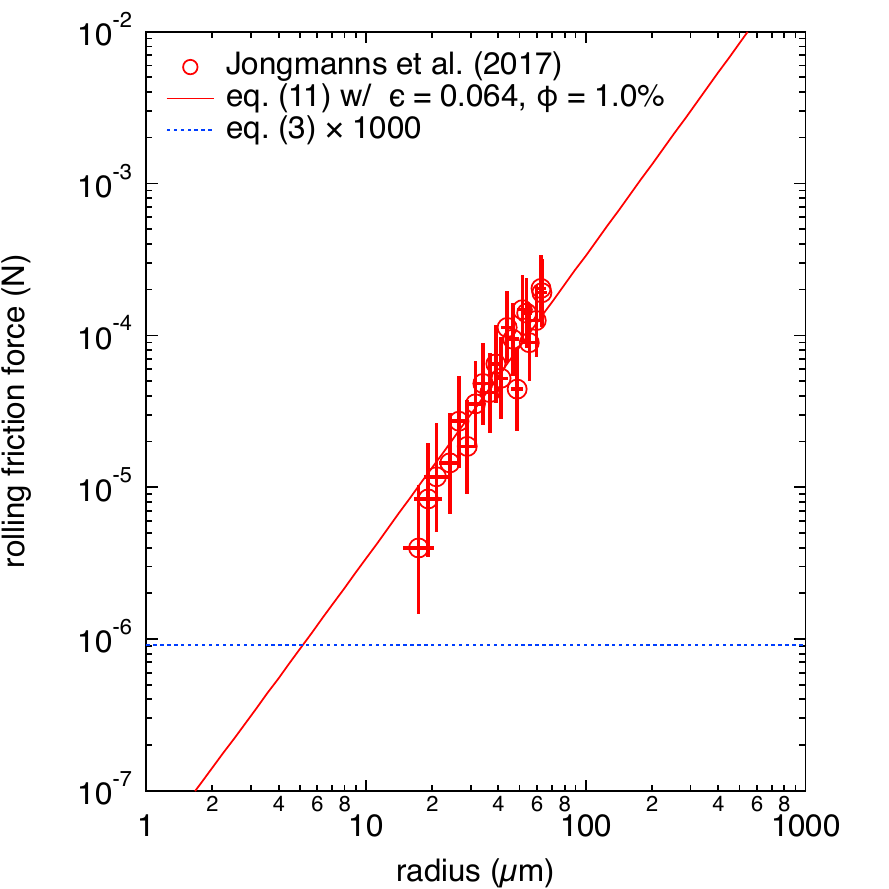}
\caption{Rolling friction forces of crystalline $\mathrm{H_2O}$ particles with radius $r_0 \approx 36~\micron$ at a temperature of $T = 213~\mathrm{K}$ as a function of particle radius $r_0$. Circles: experimental data \citep{jongmanns-et-al2017}; solid line: equation~(\ref{eq:froll_cap_full}); dotted line: $1000 \times$equation~(\ref{eq:froll_JKR}) or (\ref{eq:froll_ice}).
\label{fig:jongmanns-et-al2017}}
\end{figure}
Laboratory experiments in medium vacuum at a pressure of $5~\mathrm{Pa}$ by \citet{jongmanns-et-al2017} were designed to measure the maximum centrifugal forces that crystalline water ice particles with $r_0 = 36.4 \pm 12.1~\micron$ can resist at $T \approx 213~\mathrm{K}$.
Because a combination of equations~(\ref{eq:froll_JKR}) and (\ref{eq:fpull_JKR}) leads to 
\begin{eqnarray}
F_\mathrm{pull} & = & \frac{R}{2 \xi_\mathrm{crit} } F_\mathrm{roll} ,
\label{eq:fc-froll}
\end{eqnarray}
they attempted to derive $F_\mathrm{pull}$ from their measurements of $F_\mathrm{roll}$ in the framework of the JKR theory.
By implicitly assuming that equation~(\ref{eq:fc-froll}) applies to their experiments and the equivalence of the maximum centrifugal forces and the rolling friction forces, they claimed that the critical pull-off forces of water-ice particles are proportional to the third power of radius, namely, $F_\mathrm{pull} \propto R^3$, because of $F_\mathrm{roll} \propto R^2$ in their measurements.
Their results apparently contradict the JKR theory, which predicts $F_\mathrm{pull} \propto R^1$ in equation~(\ref{eq:fpull_JKR}) or $F_\mathrm{roll} \propto R^0$ in equation~(\ref{eq:froll_JKR}), indicating that their assumption, in other words, equation~(\ref{eq:fc-froll}) does not hold in their experiments.

In addition to their experimental results, \citet{jongmanns-et-al2017} presented numerical results of their MD simulations on sticking of two identical spherical particles at a temperature of $223.15~\mathrm{K}$ using a coarse-grained model-Water (mW) potential.
Their simulations show that the contact area $A$ of the particles is proportional to the third power of the particle radius (i.e., $A \propto r_0^{3}$) in the range of $r_0 \approx 3$--$12~\mathrm{nm}$.
This indicates that their simulations are inconsistent with the JKR theory, which predicts $A \propto r_0^{4/3}$ for the contact area between two spherical particles.
Moreover, \citet{jongmanns-et-al2017} demonstrated that pull-off forces $F_\mathrm{pull}$ on the particles is proportional to the contact area in their simulations, irrespective of particle shape, indicating $F_\mathrm{pull} \propto r_0^{3}$.
They claimed that their experimental results are consistent with the simulated results, but it is most odd that they compare their experimental results with their simulated ones in the framework of JKR theory that predicts $F_\mathrm{pull} \propto r_0$.
If the proportionality of $A \propto r_0^{3}$ is hold for $r_0 \gtrsim 12~\mathrm{nm}$, then the contact area exceeds the geometrical cross section of the particles at $r_0 \gtrsim 17~\mathrm{nm}$.
Therefore, it is physically not feasible to extend the results of their MD simulations to $10~\micron$-sized particles used in their experiments.

Their experimental results imply that the measured forces are proportional to $R^2$, which is inconsistent with the JKR theory given in equation~(\ref{eq:fslide_JKR}), but agrees with the second term of equation~(\ref{eq:froll_cap_full}).
Therefore, we may examine a possibility that the experimental results of \citet{jongmanns-et-al2017} are accounted for by lubrication rolling friction forces given in equation~(\ref{eq:froll_cap_full}).
In Fig.~\ref{fig:jongmanns-et-al2017}, we show an excellent agreement between their experimental data and equation~(\ref{eq:froll_cap_full}) with $\epsilon = 0.064$ and $\phi = 1.0\%$.
\begin{figure}
\centering\includegraphics[width=0.4\columnwidth]{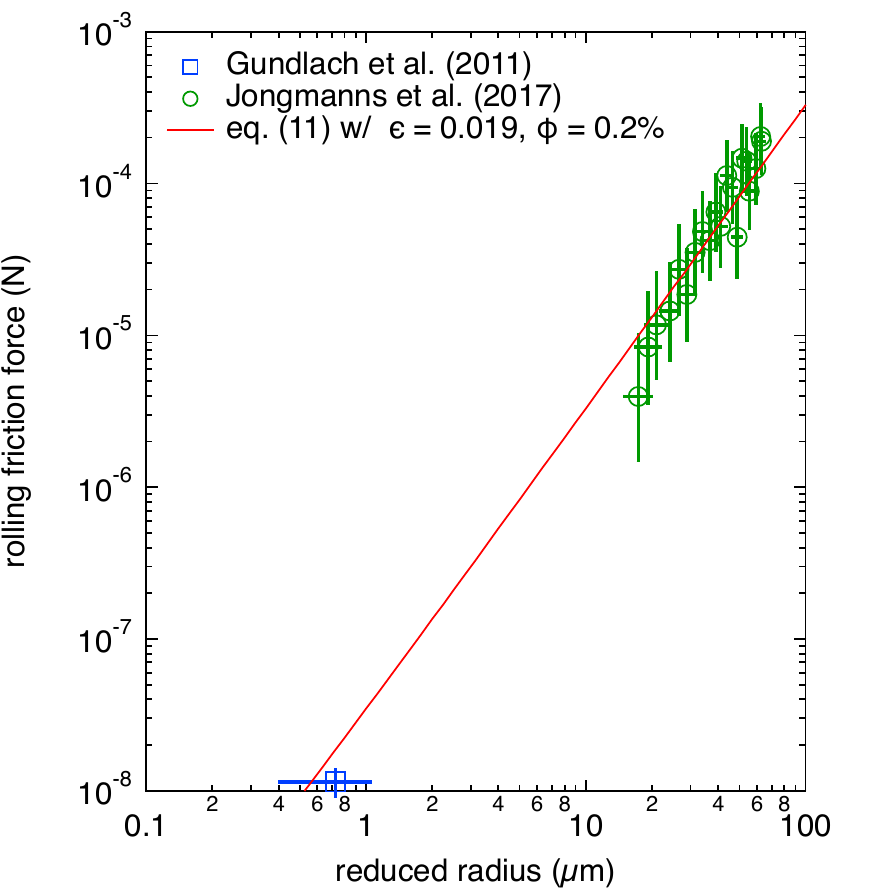}
\caption{Rolling friction forces on micrometre-sized crystalline $\mathrm{H_2O}$ particles. Open squares and circles: laboratory experiments \citep{gundlach-et-al2011,jongmanns-et-al2017}; solid line: equation~(\ref{eq:froll_cap_full}) with dissipated energy fraction $\epsilon = 0.019$ and relative humidity $\phi = 0.2\%$.
\label{fig:Froll}}
\end{figure}
While rolling friction forces were measured by \citet{gundlach-et-al2011} and \citet{jongmanns-et-al2017} at different methods and atmospheric conditions but similar temperatures, we may combine their results to compare them together to equation~(\ref{eq:froll_cap_full}).
Figure~\ref{fig:Froll} shows that both the experimental results of rolling friction forces agree with a single fitting curve of equation~(\ref{eq:froll_cap_full}) with $\epsilon = 0.019$ and $\phi = 0.2\%$.
This indicates that the results of \citet{jongmanns-et-al2017} are indeed consistent with those of \citet{gundlach-et-al2011}, provided that their rolling friction forces of water-ice particles originate from lubrication of QLLs on the surfaces of their water-ice particles.

\subsubsection{\citet{musiolik-wurm2019}}

\citet{musiolik-wurm2019} intended to measure rolling friction forces on water ice particles of $r_0 = 1.11~\mathrm{mm}$ at $150~\mathrm{Pa}$ in the range of temperatures from $180$ to $230~\mathrm{K}$.
Their results presented a temperature dependence of the friction forces with a plateau at $F_\mathrm{roll} \sim 4$--$5 \times {10}^{-4}~\mathrm{N}$ between $190$ and $220~\mathrm{K}$, which are too high to be compatible with the JKR theory given in equation~(\ref{eq:froll_JKR}) or equation~(\ref{eq:froll_ice}) if $\xi_\mathrm{crit} = 0.2~\mathrm{nm}$.
They attempted to remedy the discrepancies between their experimental results and the JKR theory by introducing asperities and temperature dependent surface energies.
In addition to friction forces, they measured pull-off forces and found the ratio of friction forces to pull-off forces being $\sim 0.1 $ for water ice particles of $R = 1.11~\mathrm{mm}$.
This is again inconsistent with the JKR theory that predicts $F_\mathrm{roll} / F_\mathrm{pull} = 3.6 \times {10}^{-7}$ for $R = 1.11~\mathrm{mm}$ from equation~(\ref{eq:fc-froll}).

\begin{figure}
\centering\includegraphics[width=0.4\columnwidth]{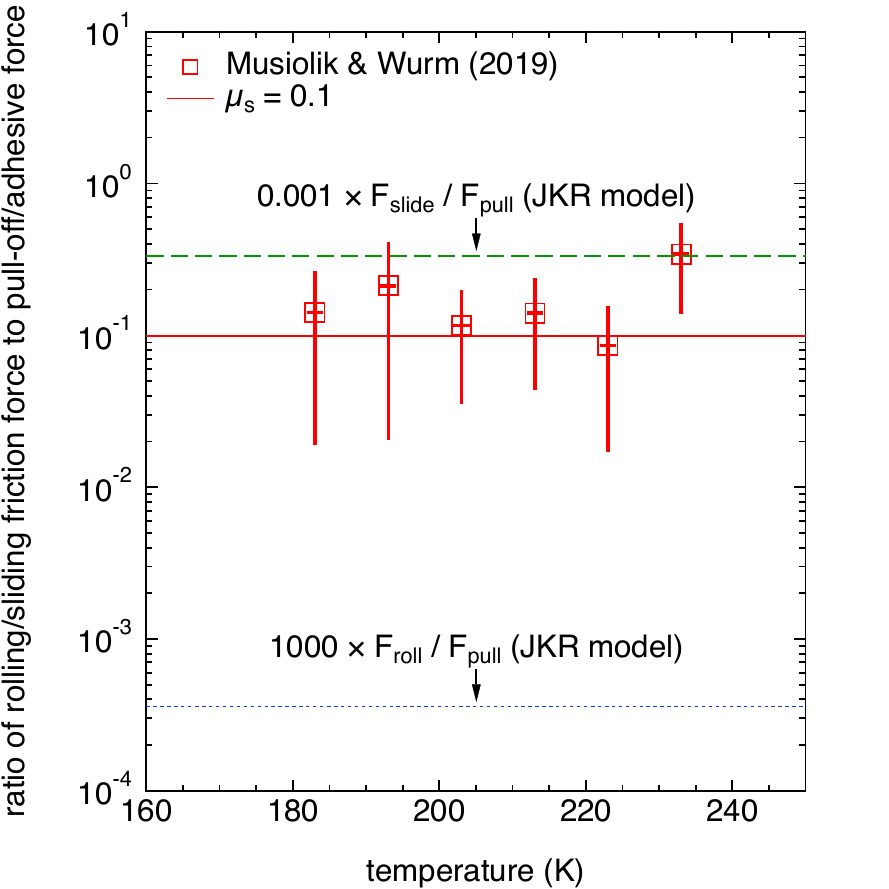}
\caption{The ratio of friction forces to pull-off forces on crystalline $\mathrm{H_2O}$ particles. Open squares: experimental data \citep{musiolik-wurm2019}; solid line: $\mu_\mathrm{s} = 0.1$ in equation~(\ref{eq:fslide_cap}); dashed line: ${10}^{-3} \times$equation~(\ref{eq:fslide_JKR})/equation~(\ref{eq:fpull_JKR}); dotted line ${10}^{3} \times$equation~(\ref{eq:froll_JKR})/equation~(\ref{eq:fpull_JKR}).
\label{fig:Fslide_Fpull}}
\end{figure}
\begin{figure}
\centering\includegraphics[width=0.4\columnwidth]{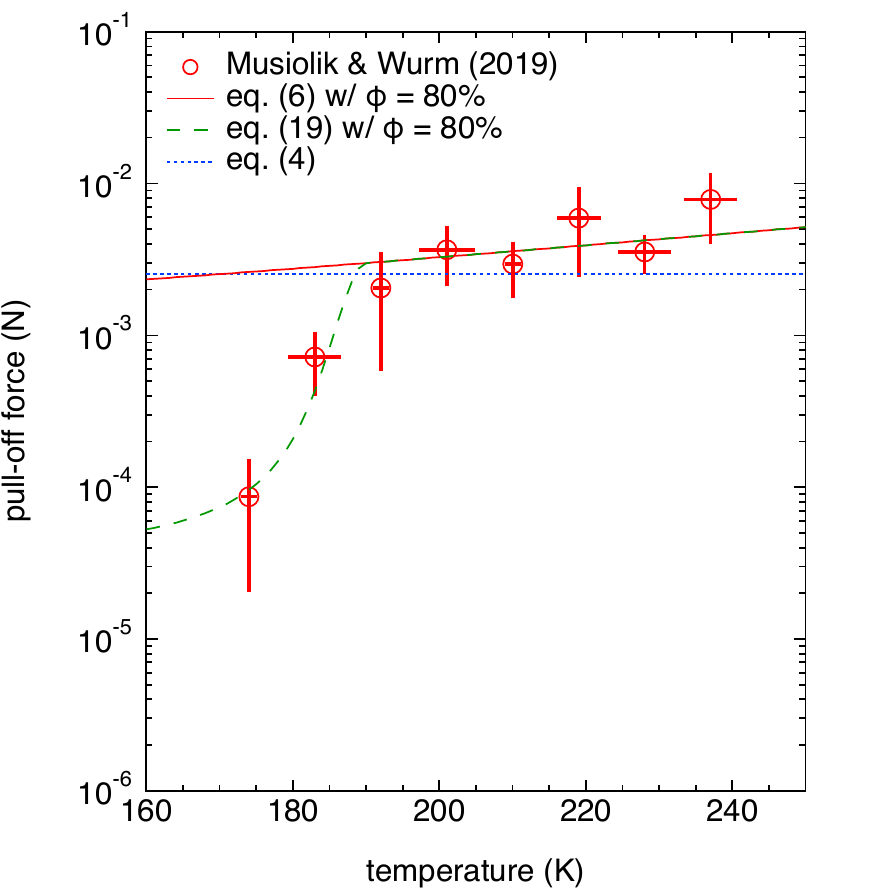}\includegraphics[width=0.4\columnwidth]{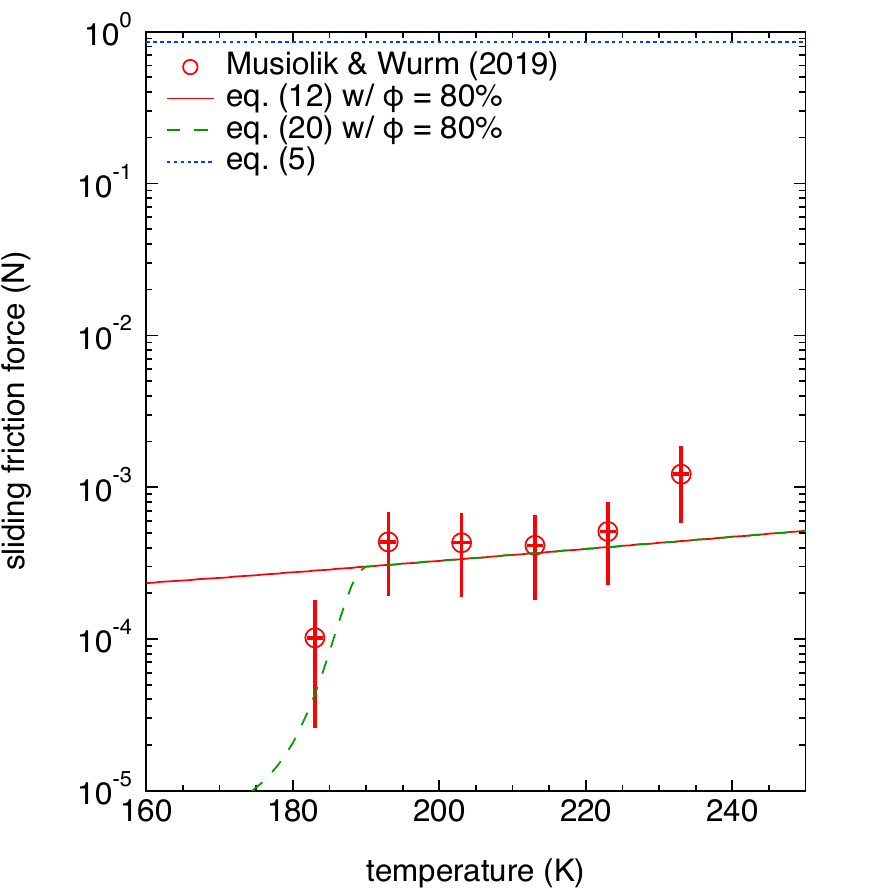}
\caption{Pull-off forces (left) and sliding friction forces (right) on crystalline $\mathrm{H_2O}$ particles. Open circles: laboratory measurements \citep{musiolik-wurm2019}; dotted lines: the JKR theory of equations~(\ref{eq:fpull_JKR}) and (\ref{eq:fslide_JKR}); solid lines: lubrication theory of equations~(\ref{eq:fadh}) and (\ref{eq:fslide_cap}) with $\phi = 80\%$ ; dashed lines: lubrication theory of equations~(\ref{eq:fpull_rough}) and (\ref{eq:froll_rough}) with $\phi = 80\%$ for rough surfaces.
\label{fig:Fadh_Fslide}}
\end{figure}

We perceive a possibility that the experimental setup of \citet{musiolik-wurm2019} was appropriate to measurements of sliding friction forces rather than rolling friction forces.
According to equation~(\ref{eq:fslide_JKR}), we may describe sliding friction forces in the JKR theory as
\begin{eqnarray}
F_\mathrm{slide} & = & 0.855~\mathrm{N} \, \left({\frac{r_0}{1.11~\mathrm{mm}}}\right)^{1/3} \left[{0.045 \left({\frac{\gamma}{0.244~\mathrm{J~m^{-2}}}}\right)^{2/3} \left({\frac{E}{7~\mathrm{GPa}}}\right)^{1/3} \left[{\frac{1}{0.9375} - \frac{0.0625}{0.9375} \left({\frac{\nu}{0.25}}\right)^2}\right]^{2/3} \left\{{\frac{4}{5} + \frac{1}{5} \left({\frac{\nu}{0.25}}\right)}\right\}^{-1}} \right.  \nonumber \\
&-& 0.252 \left({\frac{\gamma}{0.244~\mathrm{J~m^{-2}}}}\right)^{2/3} \left({\frac{E}{7~\mathrm{GPa}}}\right)^{1/3} \left[{\frac{1}{0.9375} - \frac{0.0625}{0.9375} \left({\frac{\nu}{0.25}}\right)^2}\right]^{2/3} \left({\frac{a}{0.336~\mathrm{nm}}}\right)^{-3} \left({\frac{b}{0.336~\mathrm{nm}}}\right)^{3} \nonumber \\
&+& \left. {1.207 \left({\frac{\gamma}{0.244~\mathrm{J~m^{-2}}}}\right)^{5/3} \left({\frac{E}{7~\mathrm{GPa}}}\right)^{-2/3} \left[{\frac{1}{0.9375} - \frac{0.0625}{0.9375} \left({\frac{\nu}{0.25}}\right)^2}\right]^{2/3} \left({\frac{a}{0.336~\mathrm{nm}}}\right)^{-5} \left({\frac{b}{0.336~\mathrm{nm}}}\right)^{4}}\right] . 
\label{fslide_ice}
\end{eqnarray}
Unfortunately, this far exceeds critical forces measured as rolling friction by \citet{musiolik-wurm2019}, but we shall re-investigate their measurements in terms of tribology by attributing the measured forces to sliding friction.
As shown in Fig.~\ref{fig:Fslide_Fpull}, we find that the ratios of friction forces to pull-off forces $\sim 0.1$ is in excellent harmony with the sliding friction coefficient of $\mu_\mathrm{s} = 0.1$ for water ice \citep[cf.][]{israelachvili2011}.
Consequently, we may attribute the forces measured by \citet{musiolik-wurm2019} to sliding friction by lubrication, in contrast to their interpretation of the results as rolling friction by asperities (see Fig.~\ref{fig:Fadh_Fslide}).

\begin{figure}
\centering\includegraphics[width=0.4\columnwidth]{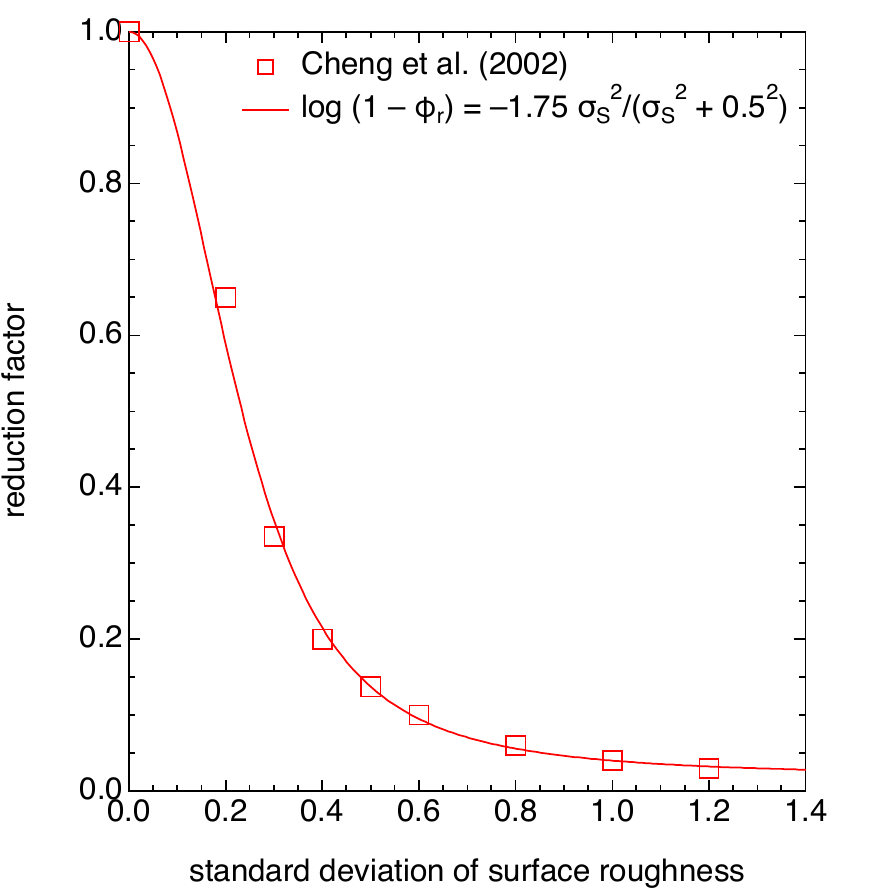}
\caption{The reduction factor $(1-\phi_\mathrm{r})$ of pull-off forces as a function of the variance of roughness heights $\sigma_\mathrm{S}$ determined by \citet{cheng-et-al2002}.
Solid line: the fitting curve given by equation~(\ref{eq:reduction_rate}).
\label{fig:reduction_rate}}
\end{figure}

Figure~\ref{fig:Fadh_Fslide} reveals that equations~(\ref{eq:fadh}) and (\ref{eq:fslide_cap}) with $\phi = 80\%$ are in harmony with pull-off and friction forces measured by \citet{musiolik-wurm2019} in the temperature range of $T > 190~\mathrm{K}$.
However, the experimental data are greatly reduced at low temperatures of $T \la 190~\mathrm{K}$, compared to equations~(\ref{eq:fadh}) and (\ref{eq:fslide_cap}), while the ratio of friction to the pull-off forces remains constant (see Fig.~\ref{fig:Fslide_Fpull}).
As shown in Fig.~\ref{fig:h_QLL}, the thickness of QLLs is smaller than the radius of water molecules in this temperature range, implying that the surface of water ice particles cannot be covered by a smooth layer of quasi-liquid water.
As a result, we should consider the surface roughness of water ice particles that reduces the cohesion between the particles according to the heights of asperities.
The reduction rate $\phi_\mathrm{r}$ of pull-off forces due to the roughness of particle surfaces computed by \citet*{cheng-et-al2002} may be approximated to
\begin{eqnarray}
\log{\left({1-\phi_\mathrm{r}}\right)} &=& -\left({\frac{7}{4}}\right) \left[{\frac{\sigma_\mathrm{S}^2}{\sigma_\mathrm{S}^2+\left({\frac{1}{2}}\right)^2}}\right] ,
\label{eq:reduction_rate}
\end{eqnarray}
where $\sigma_\mathrm{S}$ is the variance of roughness heights (see Fig.~\ref{fig:reduction_rate}).
If we assume the variance of roughness heights to be given by $\sigma_\mathrm{S} = \max{\left[{(190-T)/15,0}\right]}$ due to the appearance of asperities only at $T \la 190~\mathrm{K}$, then we have
\begin{eqnarray}
\log{\left({1-\phi_\mathrm{r}}\right)} &=& -\left({\frac{7}{4}}\right)  \left[{\max{\left({\frac{190-T}{15},0}\right)}}\right]^2 \left\{{\left[{\max{\left({\frac{190-T}{15},0}\right)}}\right]^2+\left({\frac{1}{2}}\right)^2}\right\}^{-1} .
\label{eq:reduction_rate_T}
\end{eqnarray}
Using the reduction factor $(1-\phi_\mathrm{r})$ of pull-off forces given by equation~(\ref{eq:reduction_rate_T}), we may describe the pull-off force and the sliding friction force of water ice particles as:
\begin{eqnarray}
F_\mathrm{pull} &=& \left({1 - \phi_\mathrm{r}}\right) \left[{4 \mathrm{\upi} \sigma R + \frac{\mathrm{\upi}^3}{3} \left({\frac{1-\nu^2}{E}}\right)^2 R^2 \left({\frac{\sigma}{r_\mathrm{K}}}\right)^3 + 2 \mathrm{\upi} h R \frac{\sigma}{r_\mathrm{K}}}\right] . \label{eq:fpull_rough} \\
F_\mathrm{slide} & = & \mu_\mathrm{s} \left({1 - \phi_\mathrm{r}}\right) \left[{4 \mathrm{\upi} \sigma R + \frac{\mathrm{\upi}^3}{3} \left({\frac{1-\nu^2}{E}}\right)^2 R^2 \left({\frac{\sigma}{r_\mathrm{K}}}\right)^3 + 2 \mathrm{\upi} h R \frac{\sigma}{r_\mathrm{K}}}\right]  .
\label{eq:froll_rough}
\end{eqnarray}
As demonstrated in Fig.~\ref{fig:Fadh_Fslide}, equations~(\ref{eq:fpull_rough}) and (\ref{eq:froll_rough}) (dashed lines) better reproduce both the forces measured by \citet{musiolik-wurm2019} than equations~(\ref{eq:fadh}) and (\ref{eq:fslide_cap}).
Therefore, a great reduction in the pull-off force and the sliding friction force of water ice particles in equations~(\ref{eq:fpull_rough}) and (\ref{eq:froll_rough}) most likely originates from the disappearance of QLLs at low temperatures.
\citet{musiolik-wurm2019} also attributed a reduction in the forces at temperatures below $200~\mathrm{K}$ to the disappearance of QLLs at low temperatures.
It should be, however, noted that there is a discernible difference in the concept between \citet{musiolik-wurm2019} and us as to the relation between asperities and QLLs:
\citet{musiolik-wurm2019} introduced asperities and the temperature-dependent surface energy independently to account for the imperfection of sphericity in the shape of water ice particles and the disappearance of QLLs, respectively;
Our model predicts that asperities appear when QLLs disappear, so that the importance of asperities is interconnected with the disappearance of QLLs.

\subsection{Early works in the mid-20th century}

\subsubsection{\citet{nakaya-matsumoto1954}}

\begin{figure}
\centering\includegraphics[width=0.4\columnwidth]{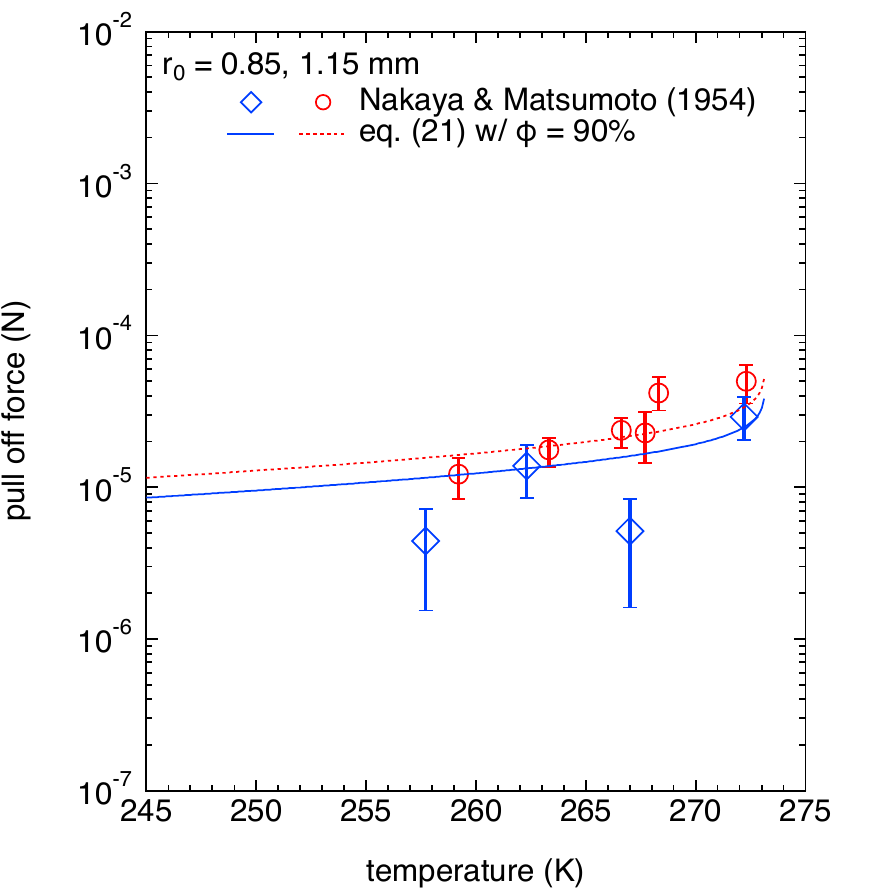}\includegraphics[width=0.4\columnwidth]{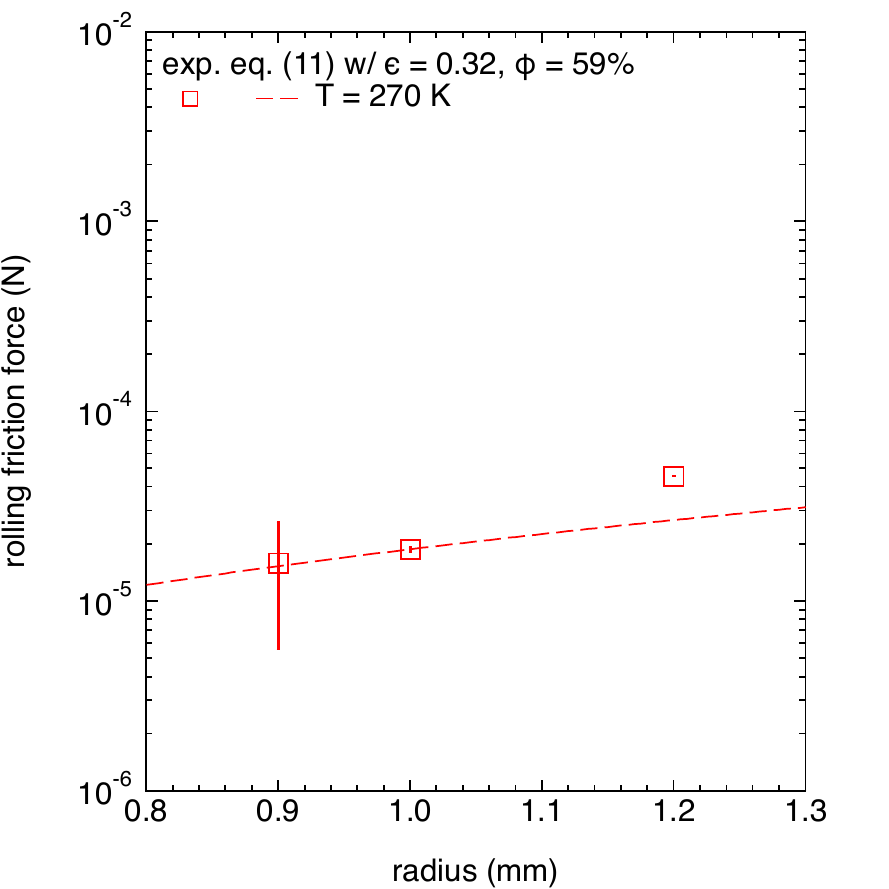}
\caption{Left: laboratory measurements of pull-off forces on millimetre-sized crystalline $\mathrm{H_2O}$ particles at atmospheric conditions by \citet{nakaya-matsumoto1954} with radius $r_0 = 0.85$ (open diamonds) and $1.15~\mathrm{mm}$ (open circles); equation~(\ref{eq:Fpull_QLLs}) with relative humidity $\phi = 90\%$ for radius $r_0 = 0.85$ (solid line) and $1.15~\mathrm{mm}$ (dotted line). Right: laboratory measurements of rolling friction forces on millimetre-sized $\mathrm{H_2O}$ particles at atmospheric conditions by \citet{nakaya-matsumoto1954} at temperature $T = 270~\mathrm{K}$ (open squares); equation~(\ref{eq:froll_cap_full}) with $\epsilon = 0.32$ and $\phi = 59\%$ (dashed line).
\label{fig:Fadh_Froll}}
\end{figure}
In the laboratory experiments by \citet{nakaya-matsumoto1954}, two touching spheres of crystalline water ice with $r_0 = 0.85$ and $1.15~\mathrm{mm}$ were suspended by thin cotton filaments at atmospheric conditions.
They pulled the particles in the direction parallel to the ground and measured an angle $\theta$ of the filaments from the normal to the ground at the time of particle separation.
This allowed them to derive the pull-off force from $F_\mathrm{pull} = m g \tan\theta$ with $m$ and $g$ being the mass of the particles and the gravitational acceleration of the earth, respectively.
Prior to particle separation, they found that the particles often rotate at an angle $\Phi$ of the filaments in the range of temperatures from $T = 266$ to $273~\mathrm{K}$.
\citet{nakaya-matsumoto1954} concluded that cohesion of water-ice spheres after the onset of rotation can be attributed to the forces due to the surface tension of QLLs.
Note that the first and the second terms in the right-hand side of equation~(\ref{eq:fadh}) vanish at zero indentation, at which two touching particles by a capillary force could easily rotate \citep{butt-et-al2010,zakerin-et-al2013}.
Since the third term in the right-hand side of equation~(\ref{eq:fadh}) remains during rotation at zero indentation, we may attribute the third term in the right-hand side of equation~(\ref{eq:fadh}) to the pull-off force $F_\mathrm{pull}$ measured by \citet{nakaya-matsumoto1954}:
\begin{eqnarray}
F_\mathrm{pull} &=& 2 \mathrm{\upi} h R \left({\frac{\sigma}{r_\mathrm{K}}}\right) .
\label{eq:Fpull_QLLs}
\end{eqnarray}
Their measurements of the angle $\Phi$ at which rotation of the particle begins also allow us to estimate the rolling friction force of crystalline water ice particles by $F_\mathrm{roll} = m g \tan\Phi$.
Figure~\ref{fig:Fadh_Froll} compares the experimental data on the pull-off force (left) and the rolling friction force (right) of crystalline water ice particles measured by \citet{nakaya-matsumoto1954} to equation~(\ref{eq:Fpull_QLLs}) with $\phi = 90\%$ and equation~(\ref{eq:froll_cap_full}) with $\epsilon = 0.32$ and $\phi = 59\%$, respectively.
Although the assumption of a constant $\phi$ value may be too crude to reproduce the experimental data, we find that equations~(\ref{eq:Fpull_QLLs}) and (\ref{eq:froll_cap_full}) with constant $\phi$ values fit the pull-off and rolling friction forces on the millimetre-sized water ice spheres measured by \citet{nakaya-matsumoto1954} within a factor of two.
Note that the pull-off forces were measured at zero indentation, implying that the radius $r_\mathrm{K}$ of curvature for meniscus is the largest, while the rolling friction forces were measured at the smallest $r_\mathrm{K}$.
Therefore, we attribute the larger $\phi$ value for the experimental data on the pull-off forces than those on the rolling friction forces to the difference in the radius $r_\mathrm{K}$ of curvature for meniscus.

\subsubsection{\citet{hosler-et-al1957} }

\begin{figure}
\centering\includegraphics[width=0.4\columnwidth]{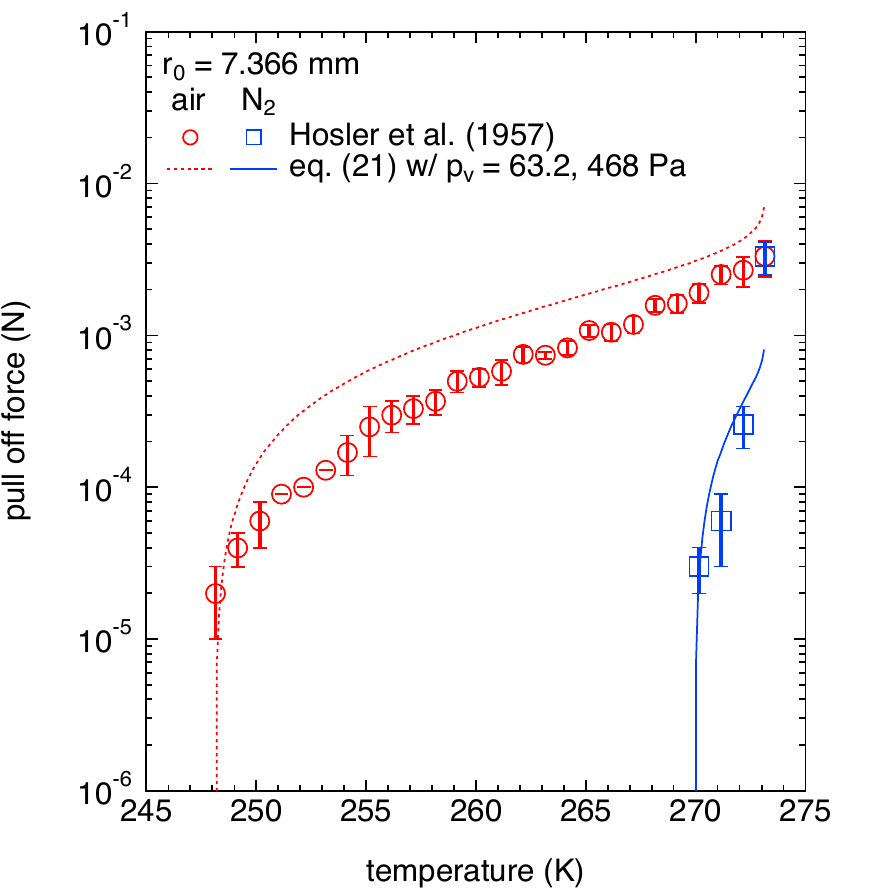}
\caption{Pull-off forces on millimetre-sized crystalline $\mathrm{H_2O}$ particles: laboratory measurements \citep{hosler-et-al1957} in air (open circles) or nitrogen vapour (open squares); equation~(\ref{eq:Fpull_QLLs}) with partial vapour pressure $p_\mathrm{v} = 63.2~\mathrm{Pa}$ (dotted line) and $468~\mathrm{Pa}$ (solid line).
\label{fig:Fpull_Hosler}}
\end{figure}
\citet{hosler-et-al1957} measured pull-off forces of spherical water-ice particles with radius $r_0 = 7.366~\mathrm{mm}$ in either air or nitrogen vapour.
Prior to their measurements, the two spheres suspended by cotton threads were brought into contact at the position of just touching, resembling the experimental setup of \citet{nakaya-matsumoto1954}.
Therefore, we assume zero indentation indicating that equation~(\ref{eq:Fpull_QLLs}) applies to their pull-off force measurements, although \citet{hosler-et-al1957} did not report whether or not the particles rotate prior to separation.
Since their experiments were carried out in a wide range of temperatures from $T = 193$ to $273~\mathrm{K}$ and the saturated vapour pressure exponentially decreases with the inverse of the temperature, we may need to consider a variation of relative humidity with the temperature.
The saturated vapour pressure $p_\mathrm{sat}$ of water molecules is given by \citep*{kimura-et-al1997}
\begin{eqnarray}
p_\mathrm{sat}(T) &=& p_\infty \exp\left({-\frac{\Delta H_\mathrm{s}}{k_\mathrm{B} T}}\right) ,
\end{eqnarray}
where $p_\infty = \displaystyle \lim_{T \to \infty} p_\mathrm{sat}(T)$ and $\Delta H_\mathrm{s}$ denotes the enthalpy of sublimation.
According to \citet{prialnik1992}, we assume the following form:
\begin{eqnarray}
p_\mathrm{sat}(T) &=& 3.56 \times {10}^{12}~\mathrm{Pa} \, \exp\left[{-\left({\frac{T}{6141.667~\mathrm{K}}}\right)^{-1}}\right] .
\end{eqnarray} 
Figure~\ref{fig:Fpull_Hosler} depicts the experimental data of \citet{hosler-et-al1957} with open circles for their measurements in air and open squares for those in nitrogen vapour.
Also plotted are the temperature variations of equation~(\ref{eq:Fpull_QLLs}) with the vapour pressure $p_\mathrm{v} = 63.2$ in air and $468~\mathrm{Pa}$ in nitrogen vapour.
Although the assumption of a constant $p_\mathrm{v}$ value may be too crude to reproduce the experimental data, equation~(\ref{eq:Fpull_QLLs}) approximates the experimental data of \citet{hosler-et-al1957} in air and nitrogen vapour within a factor of two.

\subsubsection{\citet{latham-saunders1967}}

\begin{figure}
\centering\includegraphics[width=0.4\columnwidth]{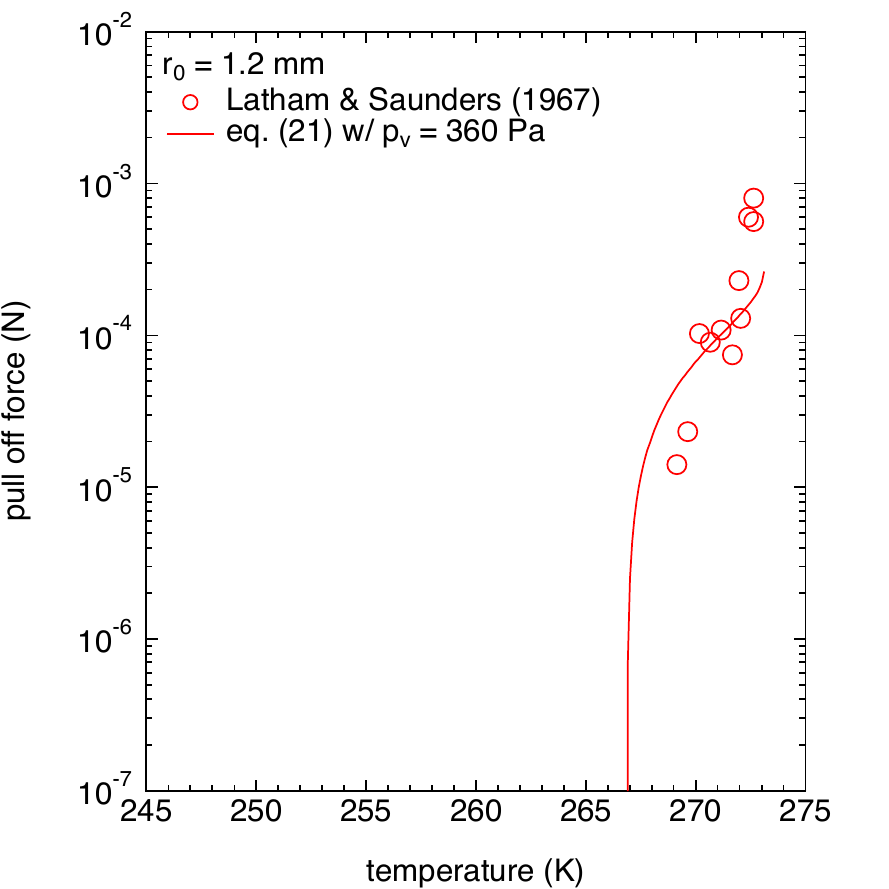}
\caption{Pull-off forces on millimetre-sized crystalline $\mathrm{H_2O}$ particles with radius $r_0 = 1.2~\mathrm{mm}$ near the melting point in dry nitrogen atmosphere. Open circles: laboratory experiments \citep{latham-saunders1967}; solid line: equation~(\ref{eq:Fpull_QLLs}) with partial vapour pressure $p_\mathrm{v} = 360~\mathrm{Pa}$.
\label{fig:Fpull_Latham}}
\end{figure}
\citet{latham-saunders1967} measured pull-off forces between two millimetre-sized ($r_0 = 1.2~\mathrm{mm}$) spheres of crystalline water ice measured near the melting point in dry nitrogen atmosphere, which is depicted in Fig.~\ref{fig:Fpull_Latham} as open circles.
They found that the pull-off forces could be one order of magnitude higher near the melting point, compared to \citet{nakaya-matsumoto1954} who used almost the same sized spheres.
\citet{latham-saunders1967} attributed the discrepancies between the pull-off forces to a difference in the vapour pressure, implying a significant role of atmospheric conditions in the pull-off force between water ice particles.
While their experimental setup does not allow spherical particles of water ice to rotate, they made gentle contact of two particles prior to their measurements of pull-off forces.
Therefore, we expect that their measurements of pull-off forces were performed for particle contacts at zero indentation, so that we can apply equation~(\ref{eq:Fpull_QLLs}) to interpret their results.
At temperatures near the melting point, we may need to take into account evaporation of water ice, which affects the relative humidity (i.e., the ratio of $p_\mathrm{v}/p_\mathrm{sat}$), since the saturated vapour pressure $p_\mathrm{sat}$ increases drastically near the melting point.
In Fig.~\ref{fig:Fpull_Latham}, we plot equation~(\ref{eq:Fpull_QLLs}) with $p_\mathrm{v} = 360~\mathrm{Pa}$ as a solid curve, supporting the idea that evaporation of water ice near the melting point affects the pull-off forces between water-ice particles.

\subsubsection{\citet{yamada-oura1970}}

\begin{figure}
\centering\includegraphics[width=0.4\columnwidth]{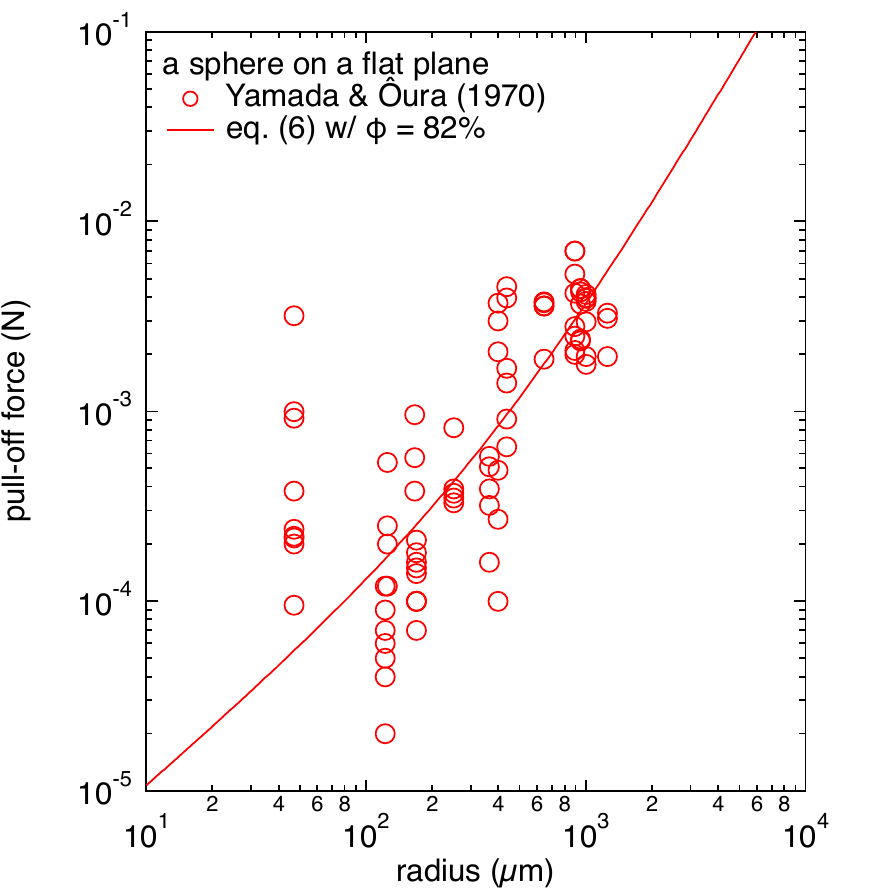}
\caption{Pull-off forces on submillimetre-sized crystalline $\mathrm{H_2O}$ spheres from a flat ice plate in air at a temperature of $T = 268.15~\mathrm{K}$. Open circles: laboratory experiments \citep{yamada-oura1970}; solid line: equation~(\ref{eq:fadh}) with relative humidity $\phi = 82\%$.
\label{fig:Fpull_Yamada}}
\end{figure}
\citet{yamada-oura1970} measured pull-off forces required to separate crystalline water ice spheres from a flat ice plate without rotation at a temperature of $T = 268.15~\mathrm{K}$.
Their experiments were conducted in air with spherical particles of radius $r_0 = 47$--$1250~\micron$, although the time-scale of sintering was not kept constant.
Since the spheres were initially brought into contact with the plate by externally applied forces, we consider that equation~(\ref{eq:fadh}) is appropriate to describe the forces measured in their experiments.
They have shown an increase in the pull-off forces with radius of spheres, which is well accounted for by equation~(\ref{eq:fadh}) with relative humidity $\phi = 82\%$ as depicted in Fig.~\ref{fig:Fpull_Yamada}.
Although the experimental data with $r_0 = 47~\micron$ deviate from equation~(\ref{eq:fadh}) with $\phi = 82\%$, their Fig.~5 suggests a tremendously huge fluctuation in their experimental conditions.

\section{Discussion}

We have shown that laboratory experiments on water ice particles are well accounted for by tribology, implying that the experimental results are to a large extent affected by the presence of QLLs on the surface of the particles.
Rare exceptions are experimental results of dynamic collision between water ice particles and aggregates of the particles, which seem to be scarcely influenced by the presence of QLLs.
Provided that the thickness of the QLLs is smaller than the maximum indentation $\delta_{\max}$ between the particles upon collision, we may assume that QLLs do not play a significant role in the outcome of collision between two particles.
On the basis of the JKR theory, the maximum indentation for two identical colliding particles is given by
\begin{eqnarray}
\delta_{\max} &=& \left[{\frac{5 \mathrm{\upi} \left({1 - \nu^2}\right) \rho v_\mathrm{imp}^{2}}{\sqrt{2} E}}\right]^{2/5} r_0 .
\label{eq:max-indentation}
\end{eqnarray}
On the assumption of $\rho = 1.0 \times {10}^{3}~\mathrm{kg~m^{-3}}$, $E = 7~\mathrm{GPa}$ and $\nu = 0.25$ for water ice, we have 
\begin{eqnarray}
\delta_{\max} &=& 43~\mathrm{nm} \, \left({\frac{v_\mathrm{imp}}{10~\mathrm{m~s^{-1}}}}\right)^{4/5} \left({\frac{\rho}{1.0 \times {10}^{3}~\mathrm{kg~m^{-3}}}}\right)^{2/5} \left({\frac{E}{7~\mathrm{GPa}}}\right)^{-2/5} \left[{\frac{1}{0.9375} - \frac{0.0625}{0.9375} \left({\frac{\nu}{0.25}}\right)^2}\right]^{2/5} \left({\frac{r_0}{1.47~\micron}}\right) . \nonumber \\
\end{eqnarray}
Since \citet{gundlach-blum2015} performed their collision experiments in the range of $v_\mathrm{imp} = 1$--$150~\mathrm{m~s^{-1}}$, we expect that the condition of $\delta_{\max} > h$ typically holds in their experiments.
This indicates that the kinetic energy of impacting particles cannot be sufficiently dissipated by QLLs and thus the effect of the QLLs on collision experiments is severely limited.

We have chosen either the relative humidity $\phi$ or the partial vapour pressure $p_\mathrm{v}$ as a free parameter to obtain reasonably well-fitting results.
While the fitting parameter does not meet a physically impossible solution such as $\phi > 100\%$ nor $p_\mathrm{v} > p_\mathrm{sat}$, we admit that the solution itself does not justify the use of a constant value for $\phi$ nor $p_\mathrm{v}$.
One could notice that the same experimental results of \citet{gundlach-et-al2011} are consistent with both $\phi = 0.2\%$ and $\phi = 1.0\%$ as shown in Figs.~\ref{fig:jongmanns-et-al2017} and \ref{fig:Froll}.
Because this clearly demonstrates that the fitting value is not a unique solution, anyone taking it at face value should exercise extreme caution.
Several formulae in tribology are introduced in this work to merely exemplify a remarkable contribution of water vapour to laboratory experiments with our fitting results.
A physically reasonable solution with a better fit to experimental results would be found, if both $\phi$ and $p_\mathrm{v}$ are treated as free parameters, although a search for the best solution is beyond the scope of this paper.
Although we have not made a thorough examination of the parameters, we have secured sufficient evidence to justify the effect of water vapour on experimental results.
Therefore, to correctly understand the physics behind experimental results with water ice, one needs an attempt to control for the effect of water vapour on the experiments.

The gas pressure in protoplanetary discs and molecular clouds is usually different from the conditions in the laboratory where the mechanical properties of water ice particles have been measured.
According to \citet{hayashi1981}, the number density, temperature and pressure of gas in the mid-plane of the solar nebula fall off, as the distance from the central star, $a$, increases:
\begin{eqnarray}
n_\mathrm{v}(a) & = & 4.1 \times {10}^{20}~\mathrm{m^{-3}} \, \left({\frac{a}{a_{\earth}}}\right)^{-11/4} ,  \\
T_\mathrm{v}(a) & = & 280~\mathrm{K} \, \left({\frac{a}{a_{\earth}}}\right)^{-1/2} \left({\frac{L_{\star}}{L_{\sun}}}\right)^{1/4} , \\
p_\mathrm{v}(a) & = & 1.6~\mathrm{Pa} \, \left({\frac{a}{a_{\earth}}}\right)^{-13/4}  \left({\frac{L_{\star}}{L_{\sun}}}\right)^{1/4} ,  
\end{eqnarray}
where $L_{\star}$ and $L_{\sun}$ denote the luminosities of the central star and the Sun, respectively, and $a_{\earth} = 1~\mathrm{au}$.
By considering the condition of $p_\mathrm{v}(a) > p_\mathrm{sat}(T_\mathrm{v})$, in which water vapour condenses into ices, one could find that water ice exists under pressure of $p_\mathrm{v} \la 0.2~\mathrm{Pa}$ beyond the snow line at $a \approx 2~\mathrm{au}$.
In the dense core of molecular clouds, we may expect a Plummer-like density profile \citep{whitworth-bate2002,lefevre-et-al2014}
\begin{eqnarray}
n_\mathrm{v}(a) & = & 3 \times {10}^{10}~\mathrm{m^{-3}} \left[{1 + \left({\frac{a}{5000~a_{\earth}}}\right)^2}\right]^{-1} ,  \\
T_\mathrm{v}(a) & = & 10~\mathrm{K} \left\{{1 + 0.007 \left[{1 + \left({\frac{a}{5000~a_{\earth}}}\right)^2}\right]}\right\} , \\
p_\mathrm{v}(a) & = & 4.1 \times {10}^{-12}~\mathrm{Pa} \left\{{\left[{1 + \left({\frac{a}{5000~a_{\earth}}}\right)^2}\right]^{-1} + 0.007 }\right\} ,
\end{eqnarray}
where $a$ denotes the distance from the centre of the core.
These parameters indicate that water vapour condenses into ices in the molecular cloud core of $a \la 2 \times {10}^{5}~\mathrm{au}$ where the condition of $3 \times {10}^{-14}~\mathrm{Pa} \la p_\mathrm{v} \la 4 \times {10}^{-12}~\mathrm{Pa}$ is achieved.
To better understand dust coagulation in protoplanetary discs and molecular clouds, experimentalists are, therefore, encouraged to conduct their laboratory experiments on water ice particles at high vacuum conditions of $p_\mathrm{v} \la 0.2~\mathrm{Pa}$ and extremely high vacuum conditions of $p_\mathrm{v} \la 4 \times {10}^{-12}~\mathrm{Pa}$, respectively.

\begin{figure}
\centering\includegraphics[width=0.4\columnwidth]{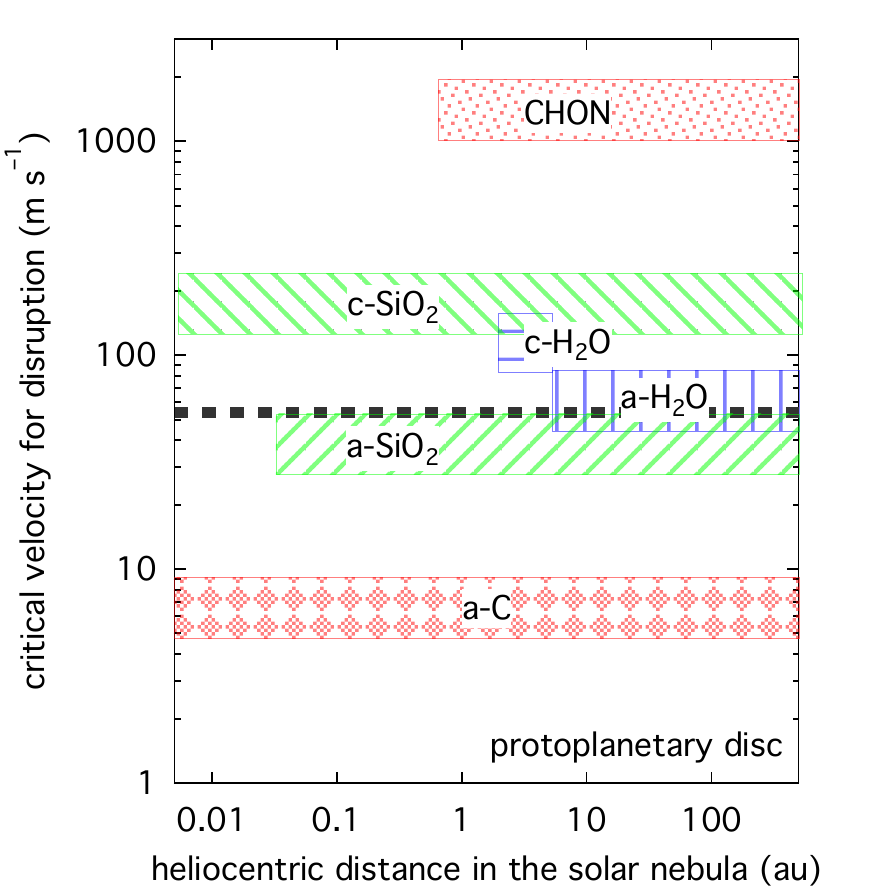}\includegraphics[width=0.4\columnwidth]{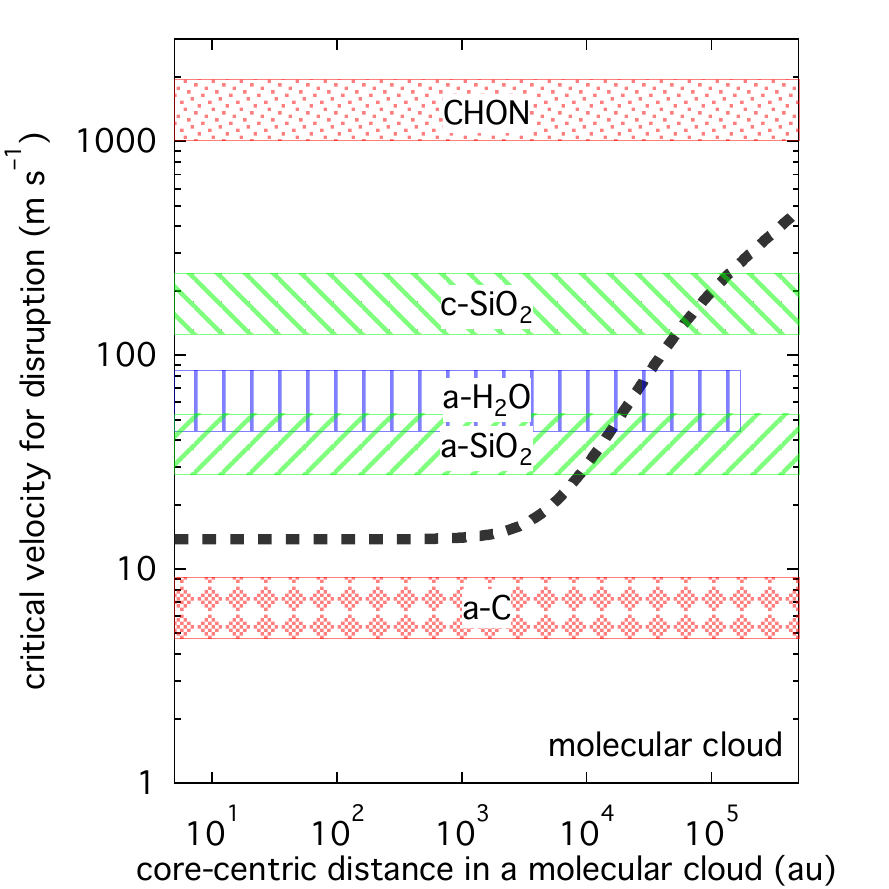}
\caption{Critical velocity for disruption of dust aggregates consisting of submicrometre-sized grains of complex organic matter (CHON), crystalline silicates (c-SiO$_2$), amorphous silicates (a-SiO$_2$), crystalline water ice (c-H$_2$O), amorphous water ice (a-H$_2$O), or amorphous carbon (a-C) in a protoplanetary disc around a solar-type star (left) and a core of a molecular cloud (right). Coagulation growth of dust aggregates is prohibited, unless the critical velocity exceeds the relative velocity of mutual collision depicted by dashed lines.
\label{fig:PPD+MC}}
\end{figure}
By considering the vacuum conditions in space, we can use the JKR theory to examine whether or not the presence of water ice on the surface of dust particles aids dust coagulation in protoplanetary discs and molecular clouds.
While previous laboratory experiments on coagulation have been carried out with crystalline water ice, H and O atoms initially condense into amorphous water ice in the outer region of protoplanetary discs and the core of molecular clouds \citep[e.g.,][]{mayer-pletzer1986}.
Therefore, it is important in principle to study dust coagulation of amorphous water ice rather than crystalline water ice, unless the temperature is high enough for amorphous materials to crystallise.
Hereafter, we consider a phase transition of amorphous materials to crystalline ones, provided that the time-scale $\tau_\mathrm{crys}$ for crystallisation is shorter than the time-scale $\tau_\mathrm{col}$ for collision (see Appendix~\ref{timescales}).
To represent the composition of dust in protoplanetary discs and molecular clouds, we consider crystalline water ice (c-H$_2$O), amorphous water ice (a-H$_2$O), crystalline silica (c-SiO$_2$), amorphous silica (a-SiO$_2$), complex organic matter (CHON) and amorphous carbon (a-C).
Table~\ref{tbl-1} lists the critical velocity of sticking, $v_\mathrm{stick}$, for particles of these compositions with $r_0 = 0.1~\micron$ and the critical velocity of disruption, $v_\mathrm{disrupt}$, for aggregates of the particles, based on the JKR theory.
Figure~\ref{fig:PPD+MC} depicts the range of $v_\mathrm{disrupt}$ in a protoplanetary disc around a solar-type star with $L_{\star} = L_{\sun}$ and $M_{\star} = M_{\sun}$ (left) and a core of a molecular cloud (right) as a function of distance from the centre of the respective system.
Also plotted as dashed lines are the maximum relative velocities of mutual collision between dust particles in protoplanetary discs and molecular clouds \citep{weidenschilling1997,draine1985}.
Note that coagulation growth of dust aggregates is prohibited, unless the critical velocity for disruption exceeds the maximum relative velocity of collision.
If we sort the compositions in order of high growth efficiency, we have CHON $>$ c-SiO$_2$ $>$ c-H$_2$O $>$ a-H$_2$O $>$ a-SiO$_2$ $>$ a-C.
This indicates that condensation of water vapour into amorphous ice or reactive accretion of amorphous ice from H and O atoms on the surface of complex organic matter or crystalline silicates reduces critical velocities against dust coagulation.
In other words, collisional growth of dust particles slows down or even ceases by the formation of water ice mantles, except for the particles composed of amorphous silica and amorphous carbon.
The growth of ice-coated dust particles is slightly eased by crystallisation of water ice in the vicinity of but outside the snow line, but the particle growth is greatly facilitated by sublimation of water ice inside the snow line for particles with the bare surface of complex organic matter or crystalline silicates.
In addition, because dust particles with an icy mantle and a refractory core are inevitably larger than bare core particles, the critical velocity of disruption for water ice aggregates diminishes according to equation~(\ref{eq:v_disrupt}), namely, $v_\mathrm{disrupt} \propto r_0^{-5/6}$.
Consequently, we conclude that water ice, in particular, in the amorphous state is not necessarily an efficient facilitator of dust coagulation in protoplanetary discs and molecular clouds.

We are aware that the JKR theory might underestimate the critical velocity of sticking for minute particles in nanometre sizes as demonstrated by recent molecular dynamics (MD) simulations, although the proportionality of $v_\mathrm{stick} \propto r_0^{-5/6}$ in equation~(\ref{eq:v_stick}) has been confirmed by MD simulations.
Using the so-called mW (``monatomic water'') potential, \citet{nietiadi-et-al2017a} performed MD simulations on mutual collision between nanoparticles ($r_0 = 15~\mathrm{nm}$) composed of amorphous water ice.
Their results revealed that collision-induced melting in the contact area of water ice particles prevents the particles from bouncing and enhances their sticking efficiencies.
According to MD simulations performed by \citet{quadery-et-al2017} and \citet*{nietiadi-et-al2017b,nietiadi-et-al2020}, silica nanoparticles of $r_0 = 1$--$25~\mathrm{nm}$ are also sticker, compared to the prediction of the JKR theory.
It should be noted that the nanoparticles used in the MD simulations are one or two orders of magnitude smaller than submicrometre-to-micrometre-sized monomers in dust aggregates, experimental results with which are in good harmony with the JKR theory \citep{kimura-et-al2015}.
On the one hand, a noticeable rise in the temperature may take place for nanoparticles upon a collision, because an increase in the temperature $\Delta T$ due to a mutual collision between two particles at critical velocities is proportional to $\Delta T \propto r_0^{-5/3}$ \citep{kalweit-drikakis2006}.
On the other hand, \citet{luo-et-al2015} found a brittle-to-ductile transition of silica glass nanofibres at room temperature as the radius of the nanofibres decreases below $r_0 = 9~\mathrm{nm}$.
This implies that the mechanical properties of nanoparticles do not necessarily represent those of submicrometre-sized dust particles in protoplanetary discs and molecular clouds.
Therefore, it is of great importance for a comprehensive study on mutual collisions of dust particles to conduct MD simulations on submicrometre-sized particles.

Lastly, we consider the possibility that the surface area of a refractory core may be significantly larger than currently thought, indicating that only a few monolayers of H$_2$O could cover the surface of dust particles at the most \citep*[see][]{potapov-et-al2020}.
This might also end up with forming icy patches or clumps on the surface of refractory cores and leaving bare refractory cores partly exposed, as experimentally observed in the early stages of mantle growth \citep[cf.][]{rosufinsen-et-al2016,marchione-et-al2019}.
On the one hand, equation~(\ref{eq:max-indentation}) suggests that the thickness of a few monolayers is substantially smaller than the maximum indentation of colliding particles of radius $r_0 \approx 0.1~\micron$ at impact velocities of $v_\mathrm{imp} \ga 10~\mathrm{m~s^{-1}}$, thus being too thin to dissipate their kinetic energies upon collision.
On the other hand, there are no apparent grounds for enhancing the dissipation of kinetic energies, on the condition that water ice is present in the form of icy clumps on the surface of bare refractory cores.
Therefore, we may conjecture that the presence of water ice has little impact on dust coagulation in protoplanetary discs and molecular clouds, as long as a thick, multilayer of H$_2$O is not permitted to form.

\section*{Acknowledgements}

We would like to thank Hiroshi Hidaka for inspiring us to begin writing this paper and Hidekazu Tanaka for a profitable discussion about the latest advances in MD simulations on mutual collisions between nanoparticles.
We are grateful to Martin McCoustra for his fruitful comments that helped us to improve the manuscript.
This work is indebted to the Grants-in-Aid for Scientific Research (KAKENHI \#19H05085) of Japan Society for the Promotion of Science (JSPS).


\section*{Data availability}

There are no new data associated with this article.

\newpage






\newpage


\appendix

\section{Characteristic time-scales}
\label{timescales}

\begin{table}
 \centering
 \caption{Thermodynamic properties.}
 \label{tbl-2}
 \begin{tabular}{lccccl}
  \hline
Composition & $\nu_\mathrm{c}$ & $E_\mathrm{a} / k_\mathrm{B}$ & $P_0$ & $\Delta H_\mathrm{s} / k_\mathrm{B}$ & Reference\\ 
 & ($\mathrm{s^{-1}}$) & ($\mathrm{K}$) & ($\mathrm{Pa}$) & ($\mathrm{K}$) & \\ 
  \hline
H$_2$O  &  $1.05 \times {10}^{13}$ & 5370 & $3.56 \times{10}^{12}$ & 6141.667 & \citet{schmitt-et-al1989,prialnik1992}\\
SiO$_2$  &  $2.00 \times {10}^{13}$ & 49190 & $3.13 \times{10}^{10}$ & 69444.67 & \citet{fabian-et-al2000,hashimoto1990} \\
CHON        & ---                      & --- & $4.25 \times {10}^{12}$ & 9477.989 & \citet{briani-et-al2013} \\
C            &  $2.85 \times {10}^{13}$ & 125800 & $9.23 \times {10}^{15}$ & 108192.2 & \citet{fischbach1963,clarke-fox1969} \\
  \hline
 \end{tabular}
\end{table}

\begin{figure}
\centering\includegraphics[width=0.4\columnwidth]{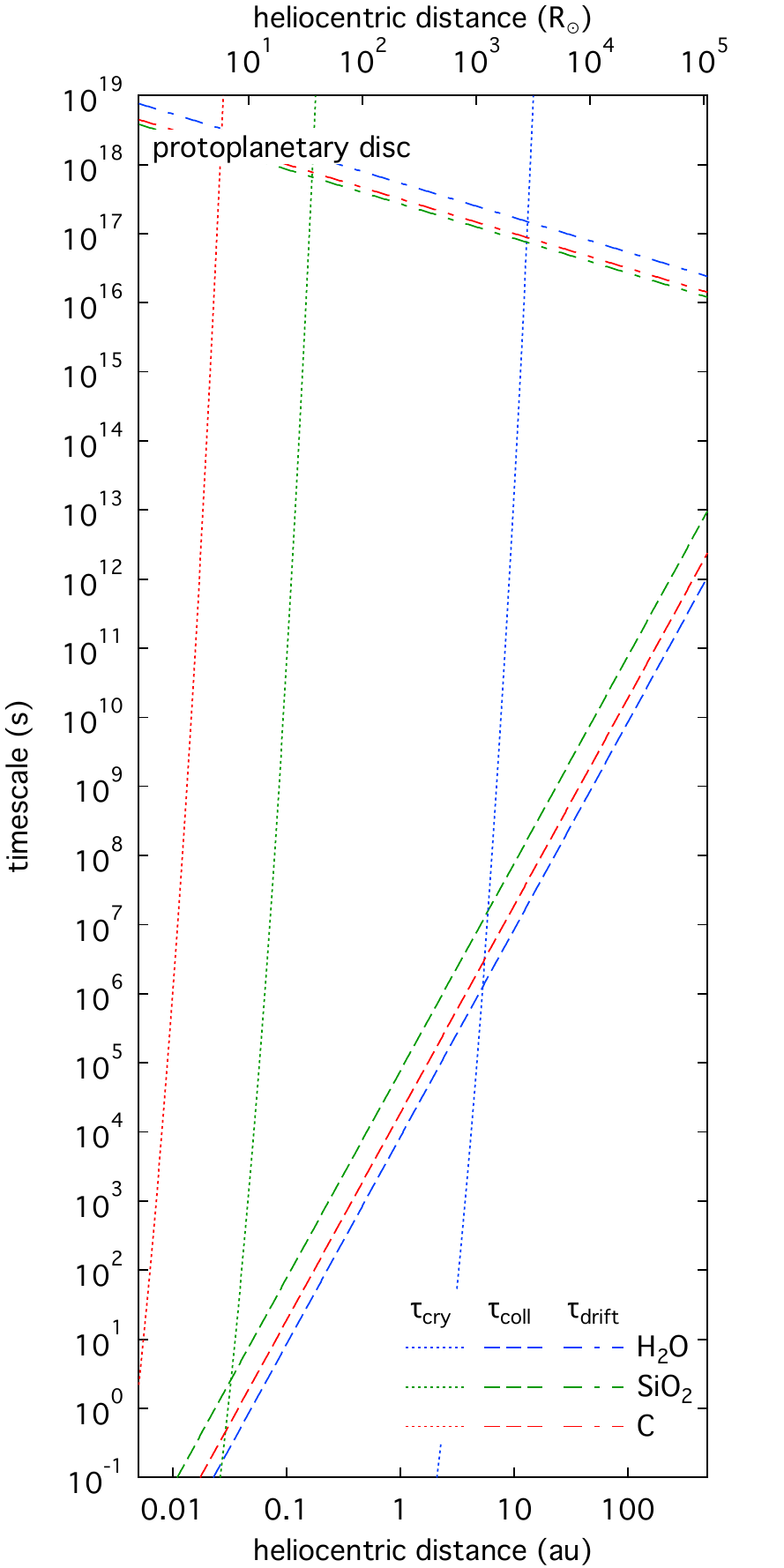}\includegraphics[width=0.4\columnwidth]{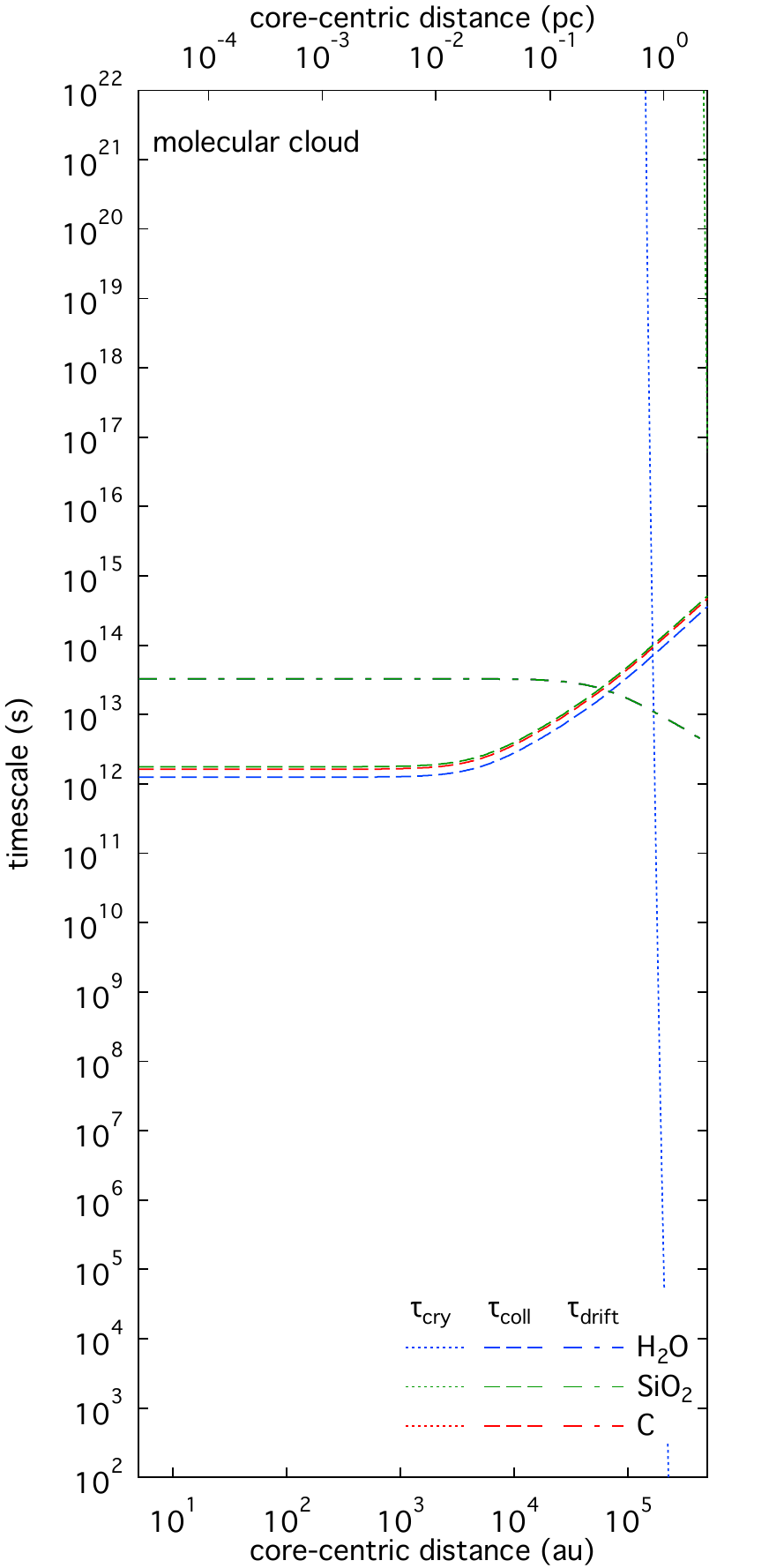}
\caption{Time-scales for crystallisation $\tau_\mathrm{crys}$, collision $\tau_\mathrm{coll}$ and drift $\tau_\mathrm{drift}$ of submicrometre-sized grains composed of silica (SiO$_2$), water ice (H$_2$O) and carbon (C) in protoplanetary discs (left) and molecular clouds (right). Coagulation growth of dust aggregates is prohibited, unless the critical velocity exceeds the relative velocity of mutual collision depicted by dashed lines.
\label{fig:timescale}}
\end{figure}

We compare characteristic time-scales for crystallisation, mutual collision and radial drift of submirometre-sized particles composed of amorphous materials in protoplanetary discs and molecular clouds.
The characteristic time-scale $\tau_\mathrm{crys}$ for crystallisation of amorphous materials is given by \citep*{lenzuni-et-al1995,kimura-et-al2002}
\begin{eqnarray}
\tau_\mathrm{crys} &=& \nu_\mathrm{c}^{-1} \exp{\left({\frac{E_\mathrm{a}}{k_\mathrm{B} T}}\right)} ,
\end{eqnarray}
where $\nu_\mathrm{c}$ is the characteristic vibrational frequency and $E_\mathrm{a}$ is the activation energy for transformation from the amorphous to the crystalline state (for the respective values, see Table~\ref{tbl-2}).
The crystallisation time-scale is compared with the characteristic time-scale $\tau_\mathrm{col}$ for mutual collision between primary particles of radius $r_0$.
The collision time-scale in protoplanetary discs is given by \citep*{brauer-et-al2008}
\begin{eqnarray}
\tau_\mathrm{col} &=& \frac{1}{\Sigma_0} \left({\frac{\mathrm{\upi} \rho^3 r_0^5 a_{\earth}^3}{54 \mu m_\mathrm{H} G M_{\star}}}\right)^{1/2} \left({\frac{a}{a_{\earth}}}\right)^{3} ,
\end{eqnarray}
where $G$ is the gravitational constant, $M_{\star}$ is the mass of the central star, $\Sigma_0$ is the surface mass density of dust particles at $a_{\earth} = 1~\mathrm{au}$ from the central star, $\mu$ is the mean molecular weight of gas ($\mu = 2.3$) and $m_\mathrm{H}$ is the mass of hydrogen \citep{hayashi1981}.
The collision time-scale in molecular clouds is given by \citep{weidenschilling-ruzmaikina1994}
\begin{eqnarray}
\tau_\mathrm{col}&=& R_\mathrm{gd} \left({\frac{r_0^2 \rho^2 l_{\max}^{2} k_\mathrm{B} T_\mathrm{v}}{72 \mathrm{\upi} v_{\max}^{6} \mu^3 m_\mathrm{H}^3 n_\mathrm{v}^2}}\right)^{1/4},
\end{eqnarray}
where $R_\mathrm{gd}$ is the gas-to-dust mass ratio ($R_\mathrm{gd} = 120$), $v_{\max}$ is the turbulent velocity ($v_{\max} = 1~\mathrm{km~s^{-1}}$) and $l_{\max}$ is the scale length in the largest eddy ($l_{\max} = 10^{13}~\mathrm{km}$) \citep[see][]{draine1985}.
We are aware that we need to take into account the the residence time of dust particles in the system, because the residence of the particles in the system is limited to a certain period of time due to a radial drift due to gas drag.
The characteristic time-scale for radial drift in protoplanetary discs is given by \citep*{adachi-et-al1976}
\begin{eqnarray}
\tau_\mathrm{drift} &=& \frac{4 \mu {m}_\mathrm{H} a}{13 k_\mathrm{B} T_\mathrm{v}} \left({\frac{G M_\star}{a}}\right)^{1/2} \frac{1 + g^2}{g} , 
\end{eqnarray}
with the dimensionless quantity $g$ being
\begin{eqnarray}
g & = & \frac{n_\mathrm{v} a}{\rho r_0} \left({\frac{8 \mathrm{\upi} \mu m_\mathrm{H} k_\mathrm{B} T_\mathrm{v}}{G M_\star / a}}\right)^{1/2}  .
\end{eqnarray}
The characteristic time-scale for drift in turbulent molecular clouds is given by \citep{boland-dejong1982}
\begin{eqnarray}
\tau_\mathrm{drift} &=& l_\mathrm{max} \left(\frac{\mathrm{\upi} \mu {m}_\mathrm{H}}{8 k_\mathrm{B} T_\mathrm{v}}\right)^{1/2} .
\end{eqnarray}
Figure~\ref{fig:timescale} depicts the above mentioned time-scales for dust particles composed of water ice (H$_2$O), silica (SiO$_2$) and carbon (C) in protoplanetary discs (left) and molecular clouds (right).
The crystallisation time-scales for SiO$_2$ and C in molecular clouds lie outside the figure, while the crystallisation time-scale for H$_2$O exceeds the time-scale for radial drift in molecular clouds, suggesting that we do not need to consider crystallisation of H$_2$O in molecular clouds.

\bsp	
\label{lastpage}
\end{document}